\documentclass[11pt,a4paper]{article}
\usepackage{psfrag}
\usepackage{graphics}
\usepackage{graphicx}
\usepackage{float}
\usepackage{amssymb,epsfig,amsmath,euscript,array}
\usepackage{cite}
\usepackage{multicol}
\usepackage{subfig}
\usepackage{fancyhdr}
\usepackage{color}
\usepackage{scalerel,stackengine}
\usepackage{pdflscape}
\usepackage{hyperref}
\usepackage{siunitx}
\makeatletter
\@addtoreset{equation}{section}
\makeatother

\usepackage[us,24hr]{datetime}

\newcounter{multieqs}

\newcommand{\be}{\begin{equation}}
\newcommand{\ee}{\end{equation}}

\newcommand{\bm}[1]{\mbox{\boldmath $#1$}}

\def\bd{\begin{document}}
\def\ed{\end{document}}
\def\nn{\nonumber}
\def\bea{\begin{eqnarray}}
\def\eea{\end{eqnarray}}
\let\bm=\bibitem
\let\la=\label

\def\npb#1#2#3{Nucl. Phys. {\bf{B#1}} #3 (#2)}
\def\plb#1#2#3{Phys. Lett. {\bf{#1B}} #3 (#2)}
\def\prl#1#2#3{Phys. Rev. Lett. {\bf{#1}} #3 (#2)}
\def\prd#1#2#3{Phys. Rev. {D \bf{#1}} #3 (#2)}
\def\cmp#1#2#3{Comm. Math. Phys. {\bf{#1}} #3 (#2)}
\def\cqg#1#2#3{Class. Quantum Grav. {\bf{#1}} #3 (#2)}
\def\nppsa#1#2#3{Nucl. Phys. B (Proc. Suppl.) {\bf{#1A}}#3 (#2)}
\def\ap#1#2#3{Ann. of Phys. {\bf{#1}} #3 (#2)}
\def\ijmp#1#2#3{Int. J. Mod. Phys. {\bf{A#1}} #3 (#2)}
\def\rmp#1#2#3{Rev. Mod. Phys. {\bf{#1}} #3 (#2)}
\def\mpla#1#2#3{Mod. Phys. Lett. {\bf A#1} #3 (#2)}
\def\jhep#1#2#3{J. High Energy Phys. {\bf #1} #3 (#2)}
\def\atmp#1#2#3{Adv. Theor. Math. Phys. {\bf #1} #3 (#2)}

\newcommand{\EQ}[1]{\begin{equation} #1 \end{equation}}
\newcommand{\AL}[1]{\begin{subequations}\begin{align} #1 \end{align}\end{subequations}}
\newcommand{\SP}[1]{\begin{equation}\begin{split} #1 \end{split}\end{equation}}
\newcommand{\ALAT}[2]{\begin{subequations}\begin{alignat}{#1} #2 \end{alignat}\end{subequations}}
\def\beq{\begin{equation}}
\def\eeq{\end{equation}}

\def\N{{\cal N}}
\def\sst{\scriptscriptstyle}
\def\thetabar{\bar\theta}
\def\Tr{{\rm Tr}}
\def\one{\mbox{1 \kern-.59em {\rm l}}}
 \def\Nh{\hat{N}}

\def\a{\alpha}      \def\da{{\dot\alpha}}
\def\b{\beta}       \def\db{{\dot\beta}}
\def\c{\gamma}  \def\G{\Gamma}  \def\cdt{\dot\gamma}
\def\d{\delta}  \def\D{\Delta}  \def\ddt{\dot\delta}
\def\e{\epsilon}        \def\vare{\varepsilon}
\def\f{\phi}    \def\F{\Phi}    \def\vvf{\f}
\def\h{\eta}
\def\k{\kappa}
\def\l{\lambda} \def\L{\Lambda}
\def\m{\mu} \def\n{\nu}
\def\o{\omega}
\def\p{\pi} \def\P{\Pi}
\def\r{\rho}
\def\s{\sigma}  \def\S{\Sigma}
\def\t{\tau}
\def\th{\theta} \def\Th{\Theta} \def\vth{\vartheta}
\def\X{\Xeta}
\def\z{\zeta}

\def\cA{{\cal A}} \def\cB{{\cal B}} \def\cC{{\cal C}}
\def\cD{{\cal D}} \def\cE{{\cal E}} \def\cF{{\cal F}}
\def\cG{{\cal G}} \def\cH{{\cal H}} \def\cI{{\cal I}}
\def\cJ{{\cal J}} \def\cK{{\cal K}} \def\cL{{\cal L}}
\def\cM{{\cal M}} \def\cN{{\cal N}} \def\cO{{\cal O}}
\def\cP{{\cal P}} \def\cQ{{\cal Q}} \def\cR{{\cal R}}
\def\cS{{\cal S}} \def\cT{{\cal T}} \def\cU{{\cal U}}
\def\cV{{\cal V}} \def\cW{{\cal W}} \def\cX{{\cal X}}
\def\cY{{\cal Y}} \def\cZ{{\cal Z}}

\def\ua{\underline{\alpha}}
\def\ub{\underline{\phantom{\alpha}}\!\!\!\beta}
\def\uc{\underline{\phantom{\alpha}}\!\!\!\gamma}
\def\um{\underline{\mu}}
\def\ud{\underline\delta}
\def\ue{\underline\epsilon}
\def\una{\underline a}\def\unA{\underline A}
\def\unb{\underline b}\def\unB{\underline B}
\def\unc{\underline c}\def\unC{\underline C}
\def\und{\underline d}\def\unD{\underline D}
\def\une{\underline e}\def\unE{\underline E}
\def\unf{\underline{\phantom{e}}\!\!\!\! f}\def\unF{\underline F}
\def\unm{\underline m}\def\unM{\underline M}
\def\unn{\underline n}\def\unN{\underline N}
\def\unp{\underline{\phantom{a}}\!\!\! p}\def\unP{\underline P}
\def\unq{\underline{\phantom{a}}\!\!\! q}
\def\unQ{\underline{\phantom{A}}\!\!\!\! Q}
\def\unH{\underline{H}}

\def\As {{A \hspace{-6.4pt} \slash}\;}
\def\bs {{b \hspace{-6.4pt} \slash}\;}
\def\Ds {{D \hspace{-6.4pt} \slash}\;}
\def\ds {{\del \hspace{-6.4pt} \slash}\;}
\def\ss {{\s \hspace{-6.4pt} \slash}\;}
\def\ks {{ k \hspace{-6.4pt} \slash}\;}
\def\ps {{p \hspace{-6.4pt} \slash}\;}
\def\pas {{{p_1} \hspace{-6.4pt} \slash}\;}
\def\pbs {{{p_2} \hspace{-6.4pt} \slash}\;}

\def\Fh{\hat{F}}
\def\Vh{\hat{V}}
\def\Xh{\hat{X}}
\def\ah{\hat{a}}
\def\xh{\hat{x}}
\def\yh{\hat{y}}
\def\ph{\hat{p}}
\def\xih{\hat{\xi}}

\def\psit{\tilde{\psi}}
\def\Psit{\tilde{\Psi}}
\def\tht{\tilde{\th}}
\def\lt{\tilde{\lambda}}
\def\At{\tilde{A}}
\def\Qt{\tilde{Q}}
\def\Rt{\tilde{R}}
\def\Nt{\tilde{N}}

\def\at{\tilde{a}}
\def\st{\tilde{s}}
\def\ft{\tilde{f}}
\def\pt{\tilde{p}}
\def\qt{\tilde{q}}
\def\vt{\tilde{v}}
\def\nt{\tilde{n}}

\def\delb{\bar{\partial}}
\def\bz{\bar{z}}
\def\bD{\bar{D}}
\def\bB{\bar{B}}

\def\bk{{\bf k}}
\def\bl{{\bf l}}
\def\bp{{\bf p}}
\def\bq{{\bf q}}
\def\br{{\bf r}}
\def\bx{{\bf x}}
\def\by{{\bf y}}
\def\bR{{\bf R}}
\def\bV{{\bf V}}

\def\d{\delta}\def\D{\Delta}\def\ddt{\dot\delta}

\def\pa{\partial} \def\del{\partial}
\def\xx{\times}
\def\uno{\mbox{1 \kern-.59em {\rm l}}}

\def\trp{^{\top}}
\def\inv{^{-1}}
\def\dag{{^{\dagger}}}
\def\pr{^{\prime}}
\def\lan{\langle}
\def\ran{\rangle}

\def\rar{\rightarrow}
\def\lar{\leftarrow}
\def\lrar{\leftrightarrow}

\newcommand{\0}{\,\!}      %this is just NOTHING!
\def\one{1\!\!1\,\,}
\def\im{\imath}
\def\jm{\jmath}

\newcommand{\tr}{\mbox{tr}}
\newcommand{\slsh}[1]{/ \!\!\!\! #1}

\def\vac{|0\rangle}
\def\lvac{\langle 0|}

\def\hlf{\frac{1}{2}}
\def\ove#1{\frac{1}{#1}}

\def\Box{\square}
\def\ZZ{\mathbb{Z}}
\def\CC#1{({\bf #1})}
\def\bcomment#1{}
\def\bfhat#1{{\bf \hat{#1}}}
\def\VEV#1{\left\langle #1\right\rangle}

\newcommand{\ex}[1]{{\rm e}^{#1}} \def\ii{{\rm i}}

\def\rr{{\rm r}} \def\rs{{\rm s}}\def\rv{{\rm v}}
\def\ri{{\rm i}}\def\rj{{\rm j}}

\newcommand{\lrbrk}[1]{\left(#1\right)}
\newcommand{\sfrac}[2]{{\textstyle\frac{#1}{#2}}}

\newcommand\equalhat{\mathrel{\stackon[1.5pt]{=}{\stretchto{%
    \scalerel*[\widthof{=}]{\wedge}{\rule{1ex}{3ex}}}{0.5ex}}}}

\font\mybb=msbm10 at 12pt
\def\bb#1{\hbox{\mybb#1}}

\font\myBB=msbm10 at 18pt
\def\BB#1{\hbox{\myBB#1}}

\newcommand{\tclr}{\textcolor}
\newcommand{\bpmat}{\begin{pmatrix}}
\newcommand{\epmat}{\end{pmatrix}}
\newcommand{\mrm}[1]{\mathrm{#1}}
\newcommand{\mrs}[1]{\scriptscriptstyle{\mathrm{#1}}}
\newcommand{\vct}[1]{\boldsymbol{#1}}
\newcommand{\hf}{\frac{1}{2}}
\newcommand{\x}{\times}
\newcommand{\pd}{\partial}
\newcommand{\dslash}{\displaystyle{\not}}
\newcommand{\ol}[1]{\overline{#1}}
\newcommand{\abs}[1]{\vert{#1}\vert}
\newcommand{\chiSqM}{\chi^2_{\mrm{min}}}
\newcommand{\chiSqMDof}{\chi^2_{\mrm{min}}/\mrm{d.o.f.}}
\newcommand{\om}{\omega}
\newcommand{\Lag}{\mathcal{L}}
\newcommand{\ord}{\mathcal{O}}
\newcommand{\eps}{\epsilon}
\newcommand{\beFrac}{\frac{1-\be}{1+\be}}
\newcommand{\beFracI}{\frac{1+\be}{1-\be}}
\newcommand{\amu}{a_{\mu}}
\newcommand{\damu}{\delta\amu}
\newcommand{\Damu}{\Delta\amu}
\newcommand{\amuUnit}{10^{-10}}
\newcommand{\mmu}{m_{\mu}}
\newcommand{\amuQED}{\amu^{\mrm{QED}}}
\newcommand{\amuEW}{\amu^{\mrm{EW}}}
\newcommand{\amuEWl}{\amu^{\mrm{EW,}\,1l}}
\newcommand{\amuEWll}{\amu^{\mrm{EW,}\,2l}}
\newcommand{\amuh}{\amu^{\mrm{had}}}
\newcommand{\amuhLO}{\amu^{\text{had, LOVP}}}
\newcommand{\amuhHO}{\amu^{\text{had, HOVP}}}
\newcommand{\amuhHOa}{\amu^{\text{had, HOVP(a)}}}
\newcommand{\amuhHOb}{\amu^{\text{had, HOVP(b)}}}
\newcommand{\amuhHOc}{\amu^{\text{had, HOVP(c)}}}
\newcommand{\amuhLbL}{\amu^{\text{had, LbL}}}
\newcommand{\ff}[3]{\mathcal{F}_{\pi^{0{#1}}\gamma^{#2}\gamma^{#3}}}
\newcommand{\alps}{\alpha_s}
\newcommand{\asmz}{\alpha_s(M_Z^2)}
\newcommand{\amz}{\alpha(M_Z^2)}
\newcommand{\aqmz}{\alpha_{\mrm{QED}}(M_Z^2)}
\newcommand{\delAlp}{\Delta\alpha}
\newcommand{\dAlpL}{\delAlp_{\mrm{lep}}}
\newcommand{\dAlpT}{\delAlp_{\mrm{top}}}
\newcommand{\dAlpH}{\delAlp_{\mrm{had}}}
\newcommand{\dAlpHF}{\dAlpH^{(5)}}
\newcommand{\dAlpHFmz}{\dAlpHF(M_Z^2)}
\newcommand{\tmin}{t_{\mrm{min}}}
\newcommand{\sTh}{s_{\mrm{th}}}
\newcommand{\eTh}{\sqrt{\sTh}}
\newcommand{\Ekmi}{E^{\,(k,m)}_i}
\newcommand{\Nkm}{N^{(k,m)}}
\newcommand{\Nkn}{N^{(k,n)}}
\newcommand{\Nexp}{N_{\mrm{exp}}}
\newcommand{\Nclu}{N_{\mrm{clu}}}
\newcommand{\Ntot}{N_{\mrm{tot}}}
\newcommand{\Rkmi}{R^{\,(k,m)}_i}
\newcommand{\Rknj}{R^{\,(k,n)}_j}
\newcommand{\dRkmi}{\mrm{d}\Rkmi}
\newcommand{\dRtkmi}{\mrm{d}\tilde{R}^{\,(k,m)}_i}
\newcommand{\BR}[2]{\mathcal{B}(#1\to #2)}
\newcommand{\decay}[2]{#1\to #2}
\newcommand{\UpsIVs}{\Upsilon(4S)}
\newcommand{\Gee}{\Gamma_{ee}}
\newcommand{\Gtot}{\Gamma_{\mrm{tot}}}
\newcommand{\ppC}{\pi^+\pi^-}
\newcommand{\ppN}{\pi^0\pi^0}
\newcommand{\pppC}{\pi^+\pi^-\pi^0}
\newcommand{\kkC}{K^+K^-}
\newcommand{\kskl}{K^0_S K^0_L}
\newcommand{\ksks}{K^0_S K^0_S}
\newcommand{\klkl}{K^0_L K^0_L}
\newcommand{\kskp}{K^0_S K^{\pm}\pi^{\mp}}
\newcommand{\eeMuMu}{e^+e^-\to\mu^+\mu^-}
\newcommand{\eeHadr}{e^+e^-\to\mrm{hadrons}}
\newcommand{\eeGhadr}{e^+e^-\to\gamma^*\to\mrm{hadrons}}
\newcommand{\tauNuHadr}{\tau\to\nu_{\tau}+\mrm{hadrons}}
\newcommand{\eeGPiPi}{e^+e^-\to\gamma^*\to\pi^+\pi^-}
\newcommand{\tauNuWNuPiPi}{\tau\to\nu_{\tau}W\to\nu_{\tau}\pi\pi^0}
\newcommand{\eeGIncl}{e^+e^-\to\gamma^*\to\mrm{all\,hadrons}}
\newcommand{\eeIncl}{e^+e^-\to\mrm{all\,hadrons}}
\newcommand{\eePiG}{e^+e^-\to\pi^0\gamma}
\newcommand{\eePiPi}{e^+e^-\to\pi^+\pi^-}
\newcommand{\eePiPiPi}{e^+e^-\to\pi^+\pi^-\pi^0}
\newcommand{\eeKK}{e^+e^-\to K^+K^-}
\newcommand{\ch}{\mrm{ch}}
\newcommand{\iso}{\mrm{iso}}
\newcommand{\noeta}{\text{no }\eta}
\newcommand{\kkr}{K\bar{K}\rho}
\newcommand{\kkp}{K\bar{K}\pi}
\newcommand{\kkpp}{K\bar{K}2\pi}
\newcommand{\kkppp}{K\bar{K}3\pi}
\newcommand{\isoAA}{(2\pi^+2\pi^-\pi^0)_{\mrm{no}\,\eta}}
\newcommand{\isoAB}{(\pi^+\pi^-3\pi^0)_{\mrm{no}\,\eta}}
\newcommand{\isoAC}{\omega(\to\mrm{npp})2\pi}
\newcommand{\isoACf}{\omega(\to\text{non-pure pionic states})2\pi}
\newcommand{\isoAD}{\eta\pi^+\pi^-}
\newcommand{\isoBA}{(2\pi^+2\pi^-2\pi^0)_{\mrm{no}\,\eta}}
\newcommand{\isoBB}{(\pi^+\pi^-4\pi^0)_{\mrm{no}\,\eta}}
\newcommand{\isoBC}{3\pi^+3\pi^-}
\newcommand{\isoBD}{\omega(\to\mrm{npp})3\pi}
\newcommand{\isoBDf}{\omega(\to\text{non-pure pionic state})3\pi}
\newcommand{\isoBE}{\eta\omega}
\newcommand{\isoEA}{\kkppp}
\newcommand{\isoEAa}{(K^+K^-\pi^+\pi^-\pi^0)_{\mrm{no}\,\eta}}
\newcommand{\isoEAb}{(K^0\bar{K}^0\pi^+\pi^-\pi^0)_{\mrm{no}\,\eta}}
\newcommand{\isoEB}{\omega(\to\mrm{npp})K\bar{K}}
\newcommand{\isoEBf}{\omega(\to\text{non-pure pionic states})K\bar{K}}
\newcommand{\isoEC}{\eta\phi}
\newcommand{\isoFA}{\eta2\pi^+2\pi^-}
\newcommand{\isoFB}{\eta\pi^+\pi^-2\pi^0}
\newcommand{\sigEEhadr}{\sigma(\eeHadr)}
\newcommand{\sigHad}{\sigma_{\mrm{had}}}
\newcommand{\sigHadB}{\sigHad^0}
\newcommand{\sigPt}{\sigma_{\mrm{pt}}}
\newcommand{\Rhad}{R_{\mrm{had}}}
\makeatletter
\def\@makefnmark{%
  \leavevmode
  \raise.9ex\hbox{\fontsize\sf@size\z@\normalfont\tiny\@thefnmark}}
\makeatother
\begin{document}

\setcounter{page}{1}
\thispagestyle{empty}
\begin{flushright}
MAN/HEP/2019/010 \\
LTH 1216 \\
KEK-TH-2165 \\
 \
\\
18th December 2019
\end{flushright}
\begin{center}

\hspace{150cm}
\
\\
\
\\
\
\\

{\LARGE{\bf The $g-2$ of charged leptons, $\alpha(M_Z^2)$ and the hyperfine splitting of muonium} \\}
\
\\
\
\
\\
\
\\
{\large Alexander Keshavarzi$^{1,2}$, Daisuke Nomura$^{3}$ and Thomas Teubner$^{4}$}
\\
\
\\
{\small \em $^1$Department of Physics and Astronomy, The University of Manchester, Manchester M13 9PL, United Kingdom} \\
{\small \em $^2$Department of Physics and Astronomy, The University of Mississippi, Mississippi 38677, U.S.} \\
{\small \em Email: alexander.keshavarzi@manchester.ac.uk} \\
{\small \em \center $^3$KEK Theory Center, Tsukuba, Ibaraki 305-0801, Japan} \\
{\small \em Email: dnomura@post.kek.jp}
{\small \em \center $^4$Department of Mathematical Sciences, University of Liverpool, Liverpool L69 3BX, United Kingdom} \\
{\small \em Email: thomas.teubner@liverpool.ac.uk}
\
\\
\
\\
\
\\
\
\\
%{\large \today } 

% ABSTRACT goes here

{\normalsize \bf Abstract}
\end{center}
Following updates in the compilation of $e^+e^-\rightarrow{\rm
  hadrons}$ data, this work presents re-evaluations of the hadronic
vacuum polarisation contributions to the anomalous magnetic moment of
the electron ($a_e$), muon ($a_\mu$) and tau lepton ($a_\tau$), to the 
ground-state hyperfine splitting of muonium and also updates the
hadronic contributions to the running of the QED coupling at the mass
scale of the $Z$ boson, $\alpha(M_Z^2)$. Combining the results for the
hadronic vacuum polarisation contributions with recent updates for the
hadronic light-by-light corrections, the electromagnetic and the weak
contributions, the deviation between the measured value of $a_\mu$
and its Standard Model prediction amounts to $\Delta a_{\mu} = 
(28.02 \pm 7.37) \times 10^{-10}$, corresponding to a muon $g-2$ 
discrepancy of $3.8\sigma$.

\newpage

\tableofcontents

\section{Introduction}\label{sec:Intro}

For the charged leptons ($l=e,\mu,\tau$), the study of their anomalous magnetic moment, $a_{l} = (g-2)_{l}/2$, continues to serve as a long-standing test of the Standard Model (SM) and as a powerful indirect search of new physics. In each case, the SM prediction of the anomalous magnetic moment is determined by summing the contributions from all sectors of the SM, such that
\beq \label{alSMeq}
a_{l}^{\rm SM}  = a_{l}^{\rm QED} + a_{l}^{\rm EW} + a_{l}^{\rm had,\,VP} + a_{l}^{\rm had,\,LbL} \, ,
\eeq
where $a_{l}^{\rm QED}$ are the QED contributions, $a_{l}^{\rm EW}$
are the (electro-)weak (EW) contributions, $a_{l}^{\rm had,\,VP}$ are the hadronic (had) vacuum polarisation (VP) contributions and $a_{l}^{\rm had,\,LbL}$ are those contributions due to hadronic light-by-light (LbL) scattering. 

The recent complete re-evaluation of the hadronic VP contributions to
$a_\mu$ preceding this work (denoted as KNT18) found the SM prediction
to be $a_{\mu}^{\rm SM}(\rm KNT18)  =  (11\ 659 \ 182.04 \pm 3.56)
\times 10^{-10}$~\cite{Keshavarzi:2018mgv}, with the uncertainty still
entirely dominated by the non-perturbative, hadronic sector. Compared
with the current experimental world average of $a_{\mu}^{\rm exp} =
(11\ 659 \ 209.1 \pm 6.3) \times
10^{-10}$~\cite{PDG2018,Bennett:2002jb,Bennett:2004pv,Bennett:2006fi},
a discrepancy of $\Delta a_{\mu} = a_{\mu}^{\rm exp} - a_{\mu}^{\rm
  SM} = (27.06 \pm 7.26)\times 10^{-10}$ was found, with the SM
prediction being $3.7\sigma$ below the experimental measurement. With
new efforts at Fermilab
(FNAL)~\cite{Grange:2015fou,Keshavarzi:2019bjn} (and later at
J-PARC~\cite{Abe:2019thb}) aiming to reduce the experimental
uncertainty by a factor of four, coupled with the ongoing efforts of
the Muon $g-2$ Theory Initiative~\cite{TGm2} to improve the
determination of the various SM contributions in conjunction with
these new measurements, it is imperative that the determination
in~\cite{Keshavarzi:2018mgv} is continuously updated and improved.

A relatively new and interesting deviation has now also arisen in the
study of the electron $g-2$. Until recently, the comparison of the
exceptionally precise measurement of $a_e^{\rm exp} = (1 \ 159 \ 652 \
180.73 \pm 0.28)  \times 10^{-12}$~\cite{Hanneke:2008tm} with the SM
prediction \allowbreak $a_{e}^{\rm SM}(\alpha_{\rm Rb})  = (1 \ 159 \
652 \ 182.032 \pm 0.720) \times 10^{-12}$~\cite{Aoyama:2017uqe} (which
updated~\cite{Aoyama:2012wj}) deviated only at the level of
$1.7\sigma$. Here, $\alpha_{\rm Rb}$ denotes that the SM prediction
has been determined using the measurement of the fine-structure
constant via rubidium (Rb) atomic
interferometry~\cite{Bouchendira:2010es}, which contributes the
dominant uncertainty to this prediction of $a_{e}^{\rm SM}$. However,
the use of a new, more precise measurement of $\alpha$ using caesium
(Cs) atomic interferometry~\cite{Parker:2018vye} results in an
estimate of $a_{e}^{\rm SM}(\alpha_{\rm Cs})  = (1 \ 159 \ 652 \
181.61 \pm 0.23) \times 10^{-12}$. This implies a deviation of $\Delta
a_{e} = a_{e}^{\rm exp} - a_{e}^{\rm SM}(\alpha_{\rm Cs}) = (-0.88 \pm
0.36)\times 10^{-12}$, corresponding to a $2.5\sigma$
difference.\footnote{Note that very recently there has been an
  independent calculation of the purely photonic five-loop
  contributions to $a_e$~\cite{Volkov:2019phy}, which gives a
  different value compared to the one in~\cite{Aoyama:2017uqe} and
  which, if adopted, would slightly change the predictions for
  $a_e^{\rm SM}$ and $\Delta a_e$.} This result has invoked much
theoretical work into the possibility of simultaneously explaining the differences in both the electron and muon sector, which must also explain the current sign difference seen between $\Delta a_{e}$ and $\Delta a_{\mu}$ (see e.g.~\cite{Crivellin:2018qmi}). Although, due to the small mass of the electron, $a_{e}^{\rm SM}$ is less sensitive to strong effects than $a_{\mu}^{\rm SM}$, the recently observed changes in the electron sector make it important that the hadronic contributions to the electron $g-2$ are also updated from the previous determination in~\cite{Nomura:2012sb} (denoted here as NT12).

Measurements of the anomalous magnetic moment of the tau lepton, $a_{\tau}^{\rm exp}$, are notoriously difficult due to the short lifetime of the $\tau$ and, as such, no direct measurement of $a_{\tau}$ has yet been achieved. Limits on $a_{\tau}^{\rm exp}$ were set by the DELPHI collaboration to be $-0.052 < a_{\tau}^{\rm exp} < 0.013$ at the 95\% confidence level~\cite{PDG2018,Abdallah:2003xd}, which is quoted in the form $a_{\tau}^{\rm exp} = -0.018(17)$ in~\cite{Abdallah:2003xd}. By standard lepton mass-scaling arguments, $a_{\tau}$ is more sensitive to heavy new physics than $a_{\mu}$ by a factor of $m_\tau^2/m_\mu^2 \sim 280$. However, the relative contributions of strong effects compared to both the electron and the muon make $a_{\tau}$ more sensitive to hadronic contributions by the same argument. The hadronic VP contributions were determined in~\cite{Eidelman:2007sb} to be $a_{\tau}^{\rm had,\,VP} = (345.1 \pm 3.9) \times 10^{-8}$, resulting (along with calculations of the various other SM contributions) in $a_{\tau}^{\rm SM} = (117 \ 721 \pm 5) \times 10^{-8}$. Although it is clear that the comparison of $\Delta a_{\tau} = a_{\tau}^{\rm exp} - a_{\tau}^{\rm SM}$ is insignificant due to the current insufficient accuracy of $a_{\tau}^{\rm exp}$, the determination of $a_{\tau}^{\rm SM}$ is an interesting undertaking and may prove useful, should experimental techniques improve to be able to better probe the anomaly of the $\tau$ lepton.

It follows that this work, denoted KNT19, will update the hadronic vacuum polarisation contributions to $a_{l} = (g-2)_{l}/2$ for all $l=e,\mu,\tau$. These are calculated utilising dispersion integrals
and the experimentally measured cross section, 
\beq \label{eq:sigma0}
\sigma^0_{{\rm had},\gamma} (s) \equiv \sigma^0(e^+e^-\rightarrow
\gamma^* \rightarrow \text{hadrons} + \gamma) \,,
\eeq
where the superscript 0 denotes the bare cross section (undressed of
all vacuum polarisation effects) and the subscript $\gamma$ indicates
the inclusion of effects from final state radiation (FSR) of (one or
more) photons
(see~\cite{Keshavarzi:2018mgv} for details). The determination of the
hadronic $R$-ratio, defined as 
\beq \label{eq:R(s)}
R(s) = \frac{\sigma^0_{{\rm had},\gamma} (s)}{\sigma_{\rm pt}(s)}
\equiv \frac{\sigma^0_{{\rm had},\gamma} (s)}{4\pi\alpha^2/(3s)}  
\eeq 
and obtained from the updated compilation of all available $e^+e^-
\rightarrow \text{hadrons}$ data, is the foundation of this endeavour. Here, $\alpha = \alpha(0)$ is the fine-structure constant. From this, the leading-order (LO) hadronic VP contributions to $a_l$ can be determined via the dispersion relation
\beq \label{eq:amu}
a_{l}^{\rm had,\,LO\,VP} =
\frac{\alpha^2}{3\pi^2}\int^{\infty}_{s_{th}} \frac{{\rm d}s}{s}
R(s)K_l(s) \,,
\eeq
where $s_{th} = m_{\pi}^2$ and $K_l(s)$ is a well-known kernel
function~\cite{Brodsky:1967sr,Lautrup:1969fr}. Expressed in the form
$\hat{K}_l(s) \equiv 3s/m_l^2 K(s)$, $\hat{K}_l(s)$ is a
monotonically-increasing function that behaves as
$\hat{K}_l(s)\rightarrow 1$ as $s\rightarrow\infty$. This behaviour
differs slightly for each lepton. In the case of the electron, the
deviation of $\hat{K}_e(s)$ from 1 is almost negligible for all $s$
and causes $a_{e}^{\rm had,\,LO\,VP} $ to be heavily dominated by the
contributions from the lowest energies~\cite{Nomura:2012sb}. For the muon, $K_\mu(s)$ behaves as $K_\mu(s) \sim m_\mu^2/(3s)$ at low energies and also accentuates the low energy domain~\cite{Hagiwara:2003da,Keshavarzi:2018mgv}, although not as heavily as for the electron. For $\hat{K}_\tau(s)$, the larger $\tau$ mass results in a functional structure that further increases the role of contributions from higher energies relative to $\hat{K}_\mu(s)$, although the role of lower energies is still prominent~\cite{Eidelman:2007sb}. At next-to-leading order (NLO), similar dispersion integrals and kernel functions exist~\cite{Krause:1996rf,Hagiwara:2003da}, allowing for $a_{l}^{\rm had,\,NLO\,VP} $ to be determined in conjunction with the LO contributions.
At NNLO, $a_{l}^{\rm had,\,NNLO\,VP} $ has been determined for $l=e,
\mu$~\cite{Kurz:2014wya}. 

In addition, the determination of the hadronic $R$-ratio is a crucial
input for two other precision observables which test the SM. First,
the hadronic contributions to the effective QED coupling
$\Delta\alpha_{\rm had}^{(5)}(q^2)$ allow for an update of this
quantity at the scale of the $Z$ boson mass, $\alpha(M_{Z}^2)$, which
hinders the accuracy of EW precision fits. Second, the hadronic VP
corrections are a non-negligible part of the ground-state hyperfine
splitting (HFS) of muonium, $\Delta\nu_{\rm Mu}$, which can be used to
determine the electron-to-muon mass ratio and, hence, the muon mass.

This paper continues, in Section~\ref{sec:DataUpdates}, with a
description of the updates in the compilation of hadronic cross data
since~\cite{Keshavarzi:2018mgv}. Section~\ref{sec:Results} details the
new results for the contributions to $a_{l}^{\rm had,\,LO\,VP}$ for
each $l=e,\mu,\tau$ (with corresponding new estimates for $a_{l}^{\rm
  SM}$), followed by updated predictions for $\alpha(M_{Z}^2)$ and
$\Delta\nu_{\rm Mu}^{\rm had, \, VP}$. Conclusions and discussions of
future prospects are given in Section~\ref{Conclusions}.

\section{Updates since the last analysis (KNT18)}\label{sec:DataUpdates}

The data combination methodology in this work is unchanged from~\cite{Keshavarzi:2018mgv} and, unless differences are explicitly stated, the cross section determination for each hadronic channel is unaltered. However, various updates with respect to the available data have been accounted for and are described in the following. As before, results for $a_{\mu}^{\rm had, \, LO \, VP}$ are quoted with their respective statistical (stat) uncertainty, systematic (sys) uncertainty, vacuum polarisation (vp) correction uncertainty and final state radiation (fsr) correction uncertainty. The total (tot) uncertainty is determined from the individual sources added in quadrature. 

\subsection{$\pi^+\pi^-$ channel}\label{chap:pipi}

The all-important $\pi^+\pi^-$ channel is modified only by the
introduction of a new radiative return measurement based on data taken
at the CLEO-c experiment between $0.3 \leq \sqrt{s} \leq 1.0$ GeV,
covering the dominant $\rho$ resonance region~\cite{Xiao:2017dqv}. The
measurement consists of two data sets: the first taken at $e^+e^-$
energies at the centre-of-mass of the $\psi(3770)$ resonance and the
second at the $\psi(4170)$ resonance. Although these measurements come
already undressed of VP effects as required by
equation~\eqref{eq:sigma0}, the undressing procedure applied
in~\cite{Xiao:2017dqv} used an outdated
routine~\cite{FJ03VP}. Therefore, in this work, the published cross
section values are redressed utilising the routine provided
in~\cite{FJ03VP} and then undressed via the KNT18 vacuum
polarisation routine,
\texttt{vp\_knt\_v3\_0}~\cite{KNT-VP,Keshavarzi:2018mgv}.\footnote{This
  routine is available for use by contacting the authors directly.}
Notably, the statistical and systematic uncertainties of the CLEO-c
data are large compared to the
KLOE~\cite{Ambrosino:2008aa,Ambrosino:2010bv,Babusci:2012rp,KLOEcombination}
and BaBar~\cite{Aubert:2009ad} measurements and, therefore, cannot
resolve the tension between the KLOE and BaBar data. In addition, in
the KNT19 data combination, the systematic uncertainties of the two CLEO-c
data sets are taken to be 100\% correlated, which further limits their influence.
 \begin{figure}[!t]
\centering
  \subfloat{%
    \includegraphics[width= 0.5\textwidth]{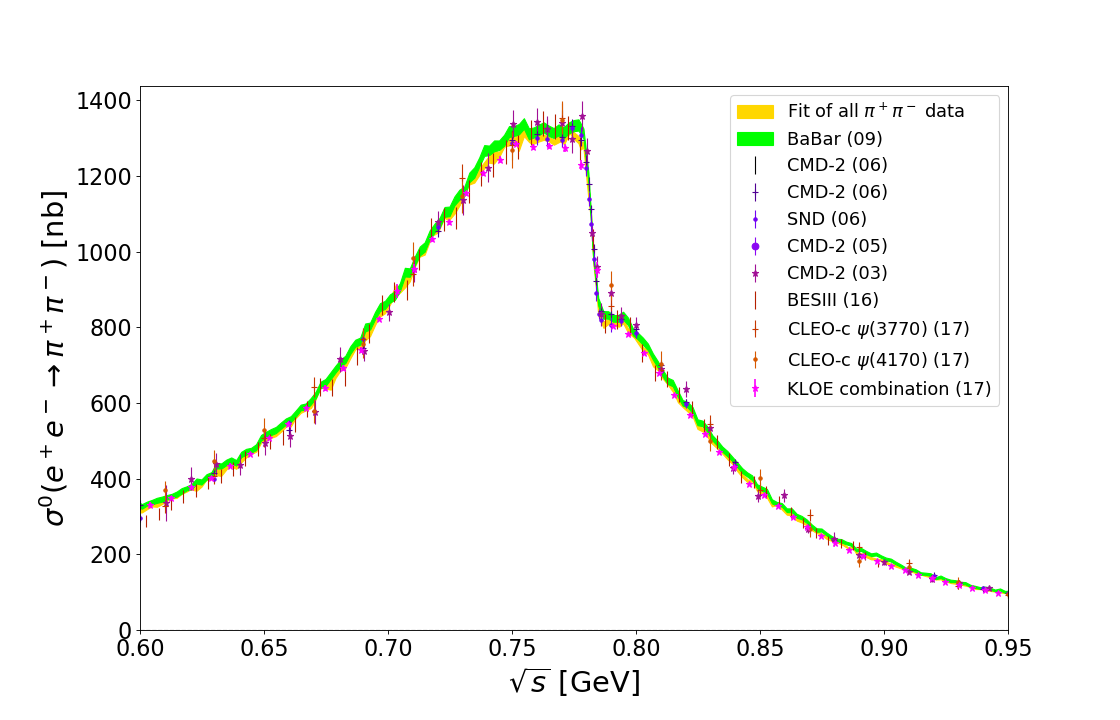}}\hfill
  \subfloat{%
    \includegraphics[width= 0.5\textwidth]{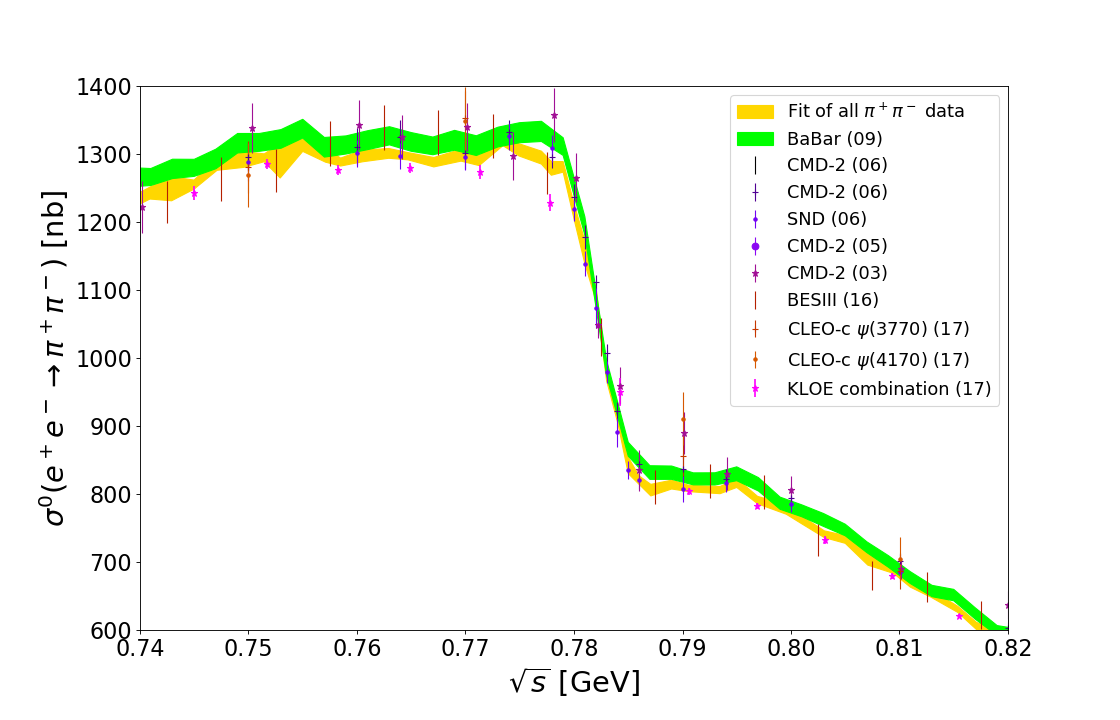}}\hfill
  \caption{\small Contributing data in the $\rho$ resonance region of the $\pi^+\pi^-$ channel plotted against the new fit of all data (left panel), with an enlargement of the $\rho$-$\omega$ interference region (right panel).}\label{fig:pipirho}
\end{figure} 
\begin{figure}[!t] 
  \centering
    \includegraphics[width=0.6\textwidth]{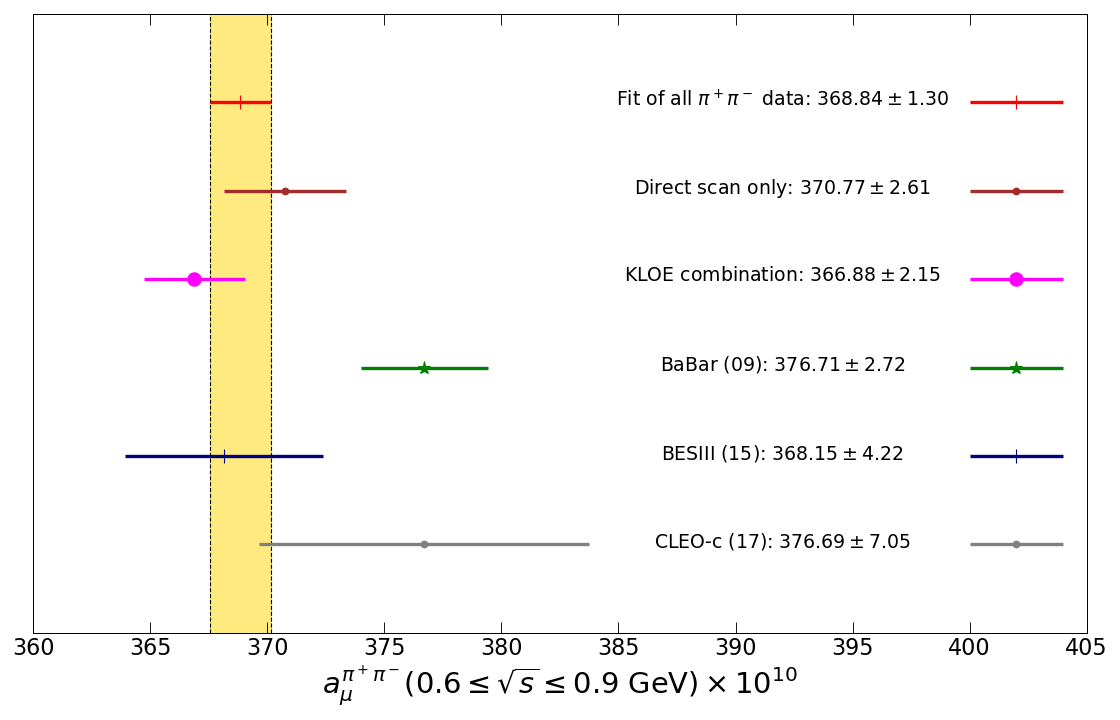}
     \caption{\small Comparison of the evaluations of $a_{\mu}^{\pi^+\pi^-}$ from the individual radiative return measurements and the combination of direct scan $\pi^+\pi^-$ measurements between $0.6 \leq \sqrt{s} \leq 0.9$ GeV.}     \label{fig:RadRetCompare}
\end{figure} 
\begin{figure}[!t] 
  \centering
    \includegraphics[width=0.7\textwidth]{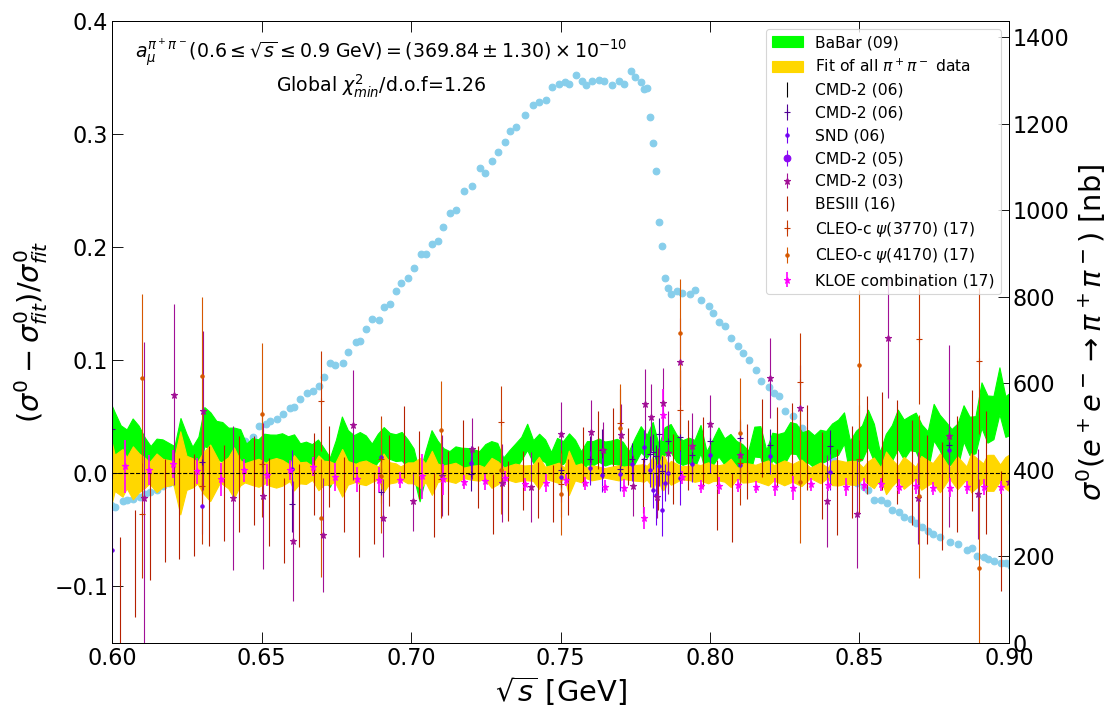}
     \caption{\small The relative difference of the radiative return
       and the most relevant of the direct scan data sets contributing to $a_{\mu}^{\pi^+\pi^-}$, and the fit of all data. For comparison, the individual sets have been normalised against the fit and have been plotted in the $\rho$ region. The green band represents the BaBar data and their errors (statistical and systematic, added in quadrature). The yellow band represents the full data combination which incorporates all correlated statistical and systematic uncertainties. However, the width of the yellow band simply displays the square root of the diagonal elements of the total output covariance matrix of the fit. }     \label{fig:RadRetFit}
\end{figure} 

The combined cross section and the dominant contributing measurements
are displayed in the $\rho$ region and magnified in the
$\rho$-$\omega$ interference region in
Figure~\ref{fig:pipirho}. Figure~\ref{fig:RadRetCompare} shows the
updated comparison of the evaluations of $a_{\mu}^{\pi^+\pi^-}$ from
the radiative return measurements and the combination of remaining
direct scan data in the vicinity of the $\rho$ resonance. Although the
new CLEO-c data are compatible with both the KLOE and BaBar
measurements, resulting in a marginal improvement in the quality of
the overall fit, as expected the combination is largely unchanged due
to the large uncertainties of the CLEO-c data. The tension between
BaBar and KLOE persists, emanated in the KNT19 combination of all
$\pi^+\pi^-$ data, which is still dominated by the three KLOE cross section measurements and their precise, highly-correlated uncertainties. This is further exemplified by Figure~\ref{fig:RadRetFit}, which clearly indicates the tension between KLOE and BaBar, and between the fit of all $\pi^+\pi^-$ data and BaBar, especially in the high-energy tail of the $\rho$ resonance. 

For the muon $g-2$, the full combination of all $\pi^+\pi^-$ data gives
\begin{align}
a_{\mu}^{\pi^+\pi^-}[0.305\leq \sqrt{s}\leq1.937\text{ GeV}] & = (503.46 \pm 1.14_{\rm stat} \pm 1.52_{\rm sys} \pm 0.06_{\rm vp} \pm 0.14_{\rm fsr}) \times 10^{-10} \nonumber
\\
\  \label{pipiFull}
& = (503.46 \pm 1.91_{\rm tot}) \times 10^{-10}\, .
\end{align}
This value is entirely consistent with~\cite{Keshavarzi:2018mgv}. The
mean value has increased by $\sim 25\%$ of the previous error, which
itself has reduced by only $\sim 3\%$. As before, tensions in the data
are accounted for in the local $\chi^2$ error inflation, increasing
the uncertainty of $a_{\mu}^{\pi^+\pi^-}$ by $\sim14\%$. This has
decreased from $\sim15\%$ in~\cite{Keshavarzi:2018mgv}, also reflected
in the slight decrease in the global $\chi^2_{\rm min}/{\rm d.o.f.}({\rm KNT18}) = 1.30$ to $\chi^2_{\rm min}/{\rm d.o.f.}({\rm KNT19}) = 1.26$ (with 625 d.o.f.).

Although the results of this work are obtained from directly integrating the combined data, detailed analyses employing constraints based on analyticity and unitarity have been performed in~\cite{Colangelo:2018mtw,Ananthanarayan:2018nyx,Ananthanarayan:2013zua,Ananthanarayan:2016mns,Davier:2019can}. These additional constraints have the potential to improve the determination of the two-pion cross section and to possibly reduce the error, especially at low energies where limited data are available. The results obtained in these works are, overall, largely compatible with the determination of this analysis, but lead to slightly larger results for $a_{\mu}^{\rm had, \, LO \, VP}$ in the energy range $\sqrt{s}<0.6$ GeV. A detailed comparison with these values is beyond the scope of this work, but will be presented as part of the studies of the Muon $g-2$ Theory Initiative~\cite{TGm2}.

\subsection{$\pi^+\pi^-\pi^0$ channel}\label{sec:3pi}

\begin{figure}[!t] 
  \centering
  \subfloat[]{%
    \includegraphics[width= 0.5\textwidth]{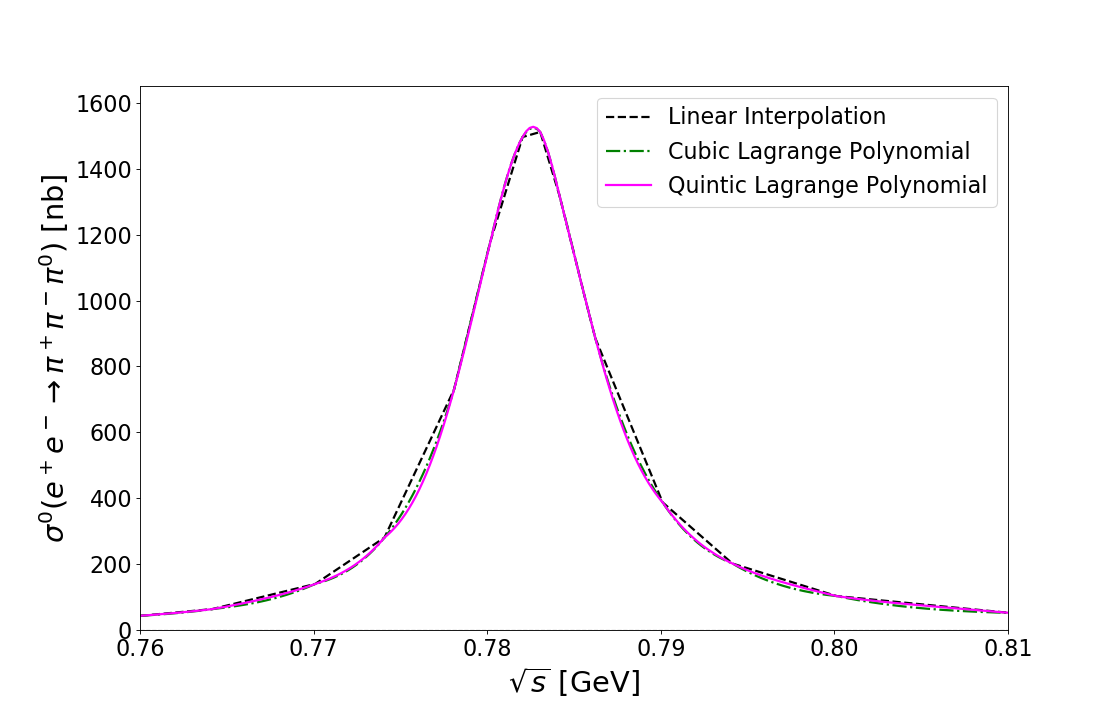}}\hfill
  \subfloat[]{%
    \includegraphics[width= 0.5\textwidth]{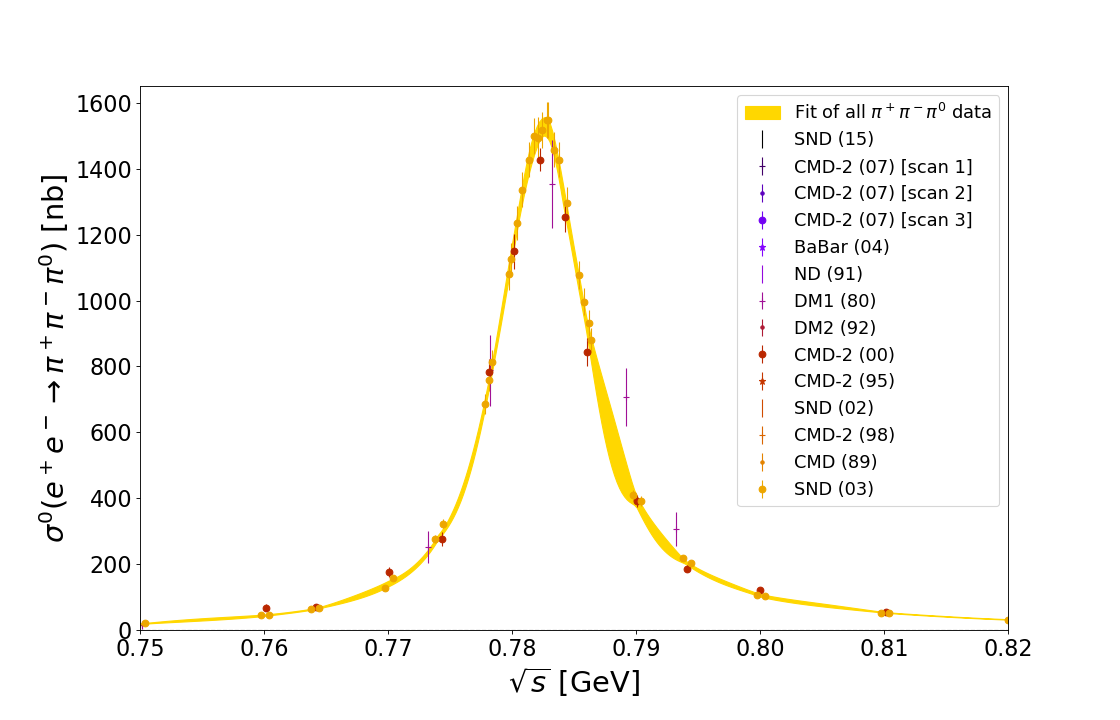}}\hfill
     \caption{\small The cross section
       $\sigma^{0}(e^+e^-\rightarrow\pi^+\pi^-\pi^0)$ in the region of
       the narrow $\omega$ resonance. In
       Figure~\ref{fig:3pi_omega}(a), the black dashed line, green
       dashed-dotted line and pink solid line show the linear, cubic and quintic interpolation between clusters, respectively.} \label{fig:3pi_omega}
\end{figure}

A recent study of the three-pion contribution to the hadronic vacuum
polarisation based on a global fit function using analyticity and
unitarity constraints~\cite{Hoferichter:2019gzf} highlighted major
differences arising in various determinations of
$a_\mu^{\pi^+\pi^-\pi^0}$. These were attributed to the choice of
cross section interpolation used in the prominent $\omega$ resonance
region when integrating the data. Due to a lack of data and a
(relatively) wide-binning in the narrow $\omega$ resonance itself, the
trapezoidal rule integration used
in~\cite{Hagiwara:2003da,Hagiwara:2006jt,Hagiwara:2011af,Keshavarzi:2018mgv},
while consistent with the direct data integration procedure utilised in these works, led to a value of $a_\mu^{\pi^+\pi^-\pi^0}$
in~\cite{Keshavarzi:2018mgv} larger than found
in~\cite{Hoferichter:2019gzf,Davier:2019can}. In order to address this
issue in this work, the clusters and covariance matrix elements
corresponding to the fitted $\omega$ resonance alone have been
interpolated to a 0.2 MeV binning using a quintic polynomial. The
newly finer-binned resonance, along with the entire $\pi^+\pi^-\pi^0$
cross section, are then integrated using the trapezoidal rule integral
to ensure consistency with the general KNT data combination procedure
applied to all other channels. This results in an improved
estimate of 
\begin{align}
a_{\mu}^{\pi^+\pi^-\pi^0}[0.66\leq \sqrt{s}\leq1.937 \text{ GeV}] & = (46.73 \pm 0.32_{\rm stat} \pm 0.74_{\rm sys} \pm 0.12_{\rm vp} \pm 0.47_{\rm fsr}) \times 10^{-10} \nonumber
\\
\
& = (46.73 \pm 0. 94_{\rm tot}) \times 10^{-10} \, ,
\end{align}
compared to $a_{\mu}^{\pi^+\pi^-\pi^0}({\rm KNT18}) = (47.79 \pm 0.89) \times 10^{-10}$
in~\cite{Keshavarzi:2018mgv}. Figure~\ref{fig:3pi_omega}(a) shows an
enlargement of the $\omega$ resonance region, where the comparison
between the previously used trapezoidal rule integral (black dashed
line), a cubic polynomial interpolation (dashed-dotted green line) and
the quintic polynomial (solid pink line) interpolation are visible,
highlighting the improvement that this change has made.\footnote{Should new data be released that better describe the shape of the $\omega$ resonance in this channel, then the higher-population
  of data may render this higher-order polynomial interpolation
  unnecessary and the trapezoidal integral over the available data may be sufficient.} 
It can also be seen here that whilst the linear interpolation clearly overestimates the resonance in the tails, the cubic interpolation seemingly underestimates and overestimates the cross section in various places in the tail, hence the choice of the quintic polynomial. The resulting KNT19 determination of the $\omega$ resonance in the $\pi^+\pi^-\pi^0$ channel and all contributing data are shown in Figure~\ref{fig:3pi_omega}(b).

\subsection{Other channels}\label{sec:OtherChannels}

\begin{figure}[!t]
\centering
 \subfloat[$\sigma^{0}(e^+e^-\rightarrow \pi\gamma)$]{%
  \includegraphics[width= 0.33\textwidth]{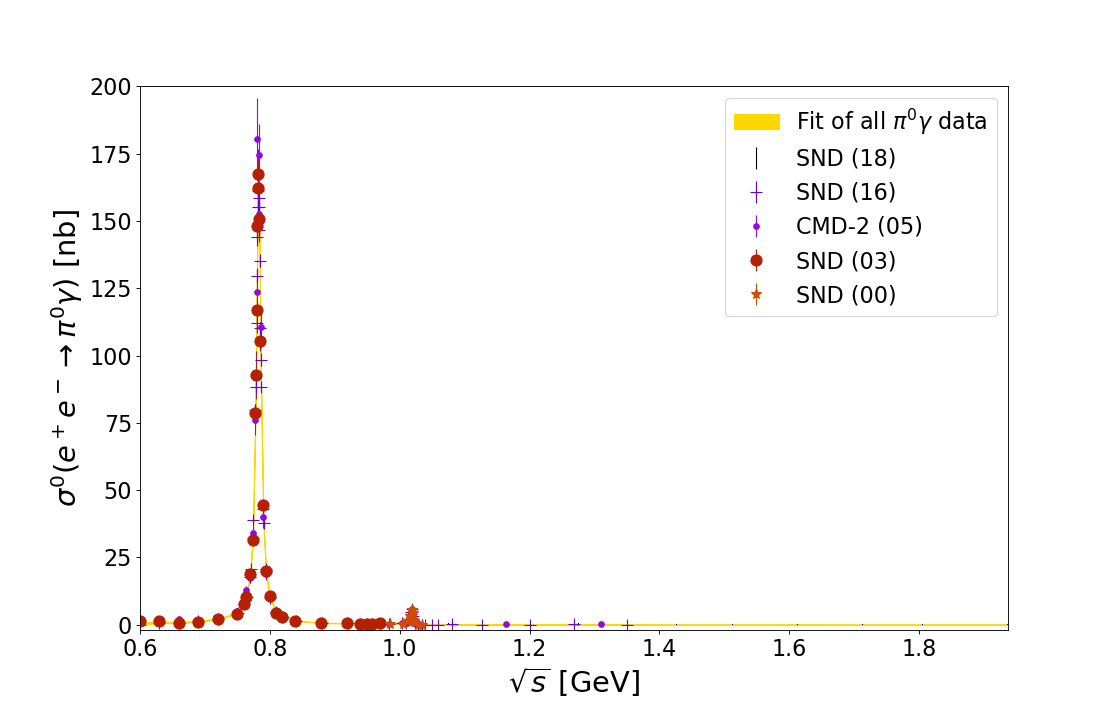}}\hfill
  \subfloat[$\sigma^{0}(e^+e^-\rightarrow \pi^+\pi^-\eta)$]{%
    \includegraphics[width= 0.33\textwidth]{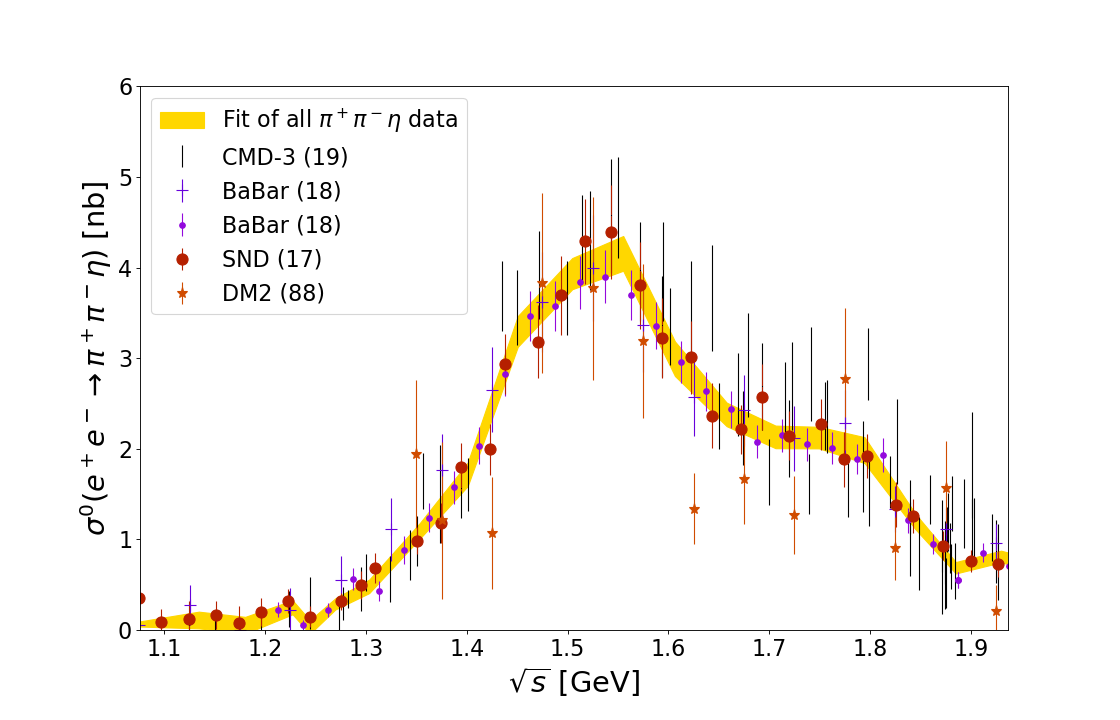}}\hfill
  \subfloat[$\sigma^{0}\big(e^+e^-\rightarrow (\pi^+\pi^-\pi^0\eta)_{{\rm no}\, \omega}\big)$]{%
    \includegraphics[width= 0.33\textwidth]{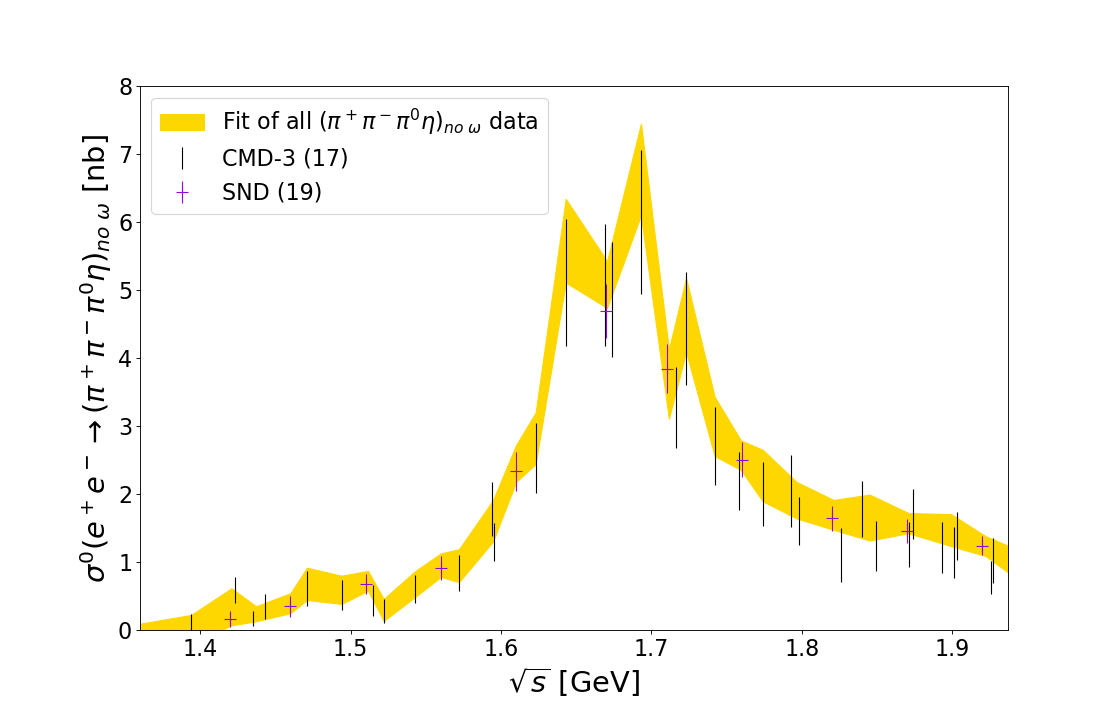}}\hfill
  \subfloat[$\sigma^{0}(e^+e^-\rightarrow \eta\omega)$]{%
    \includegraphics[width= 0.33\textwidth]{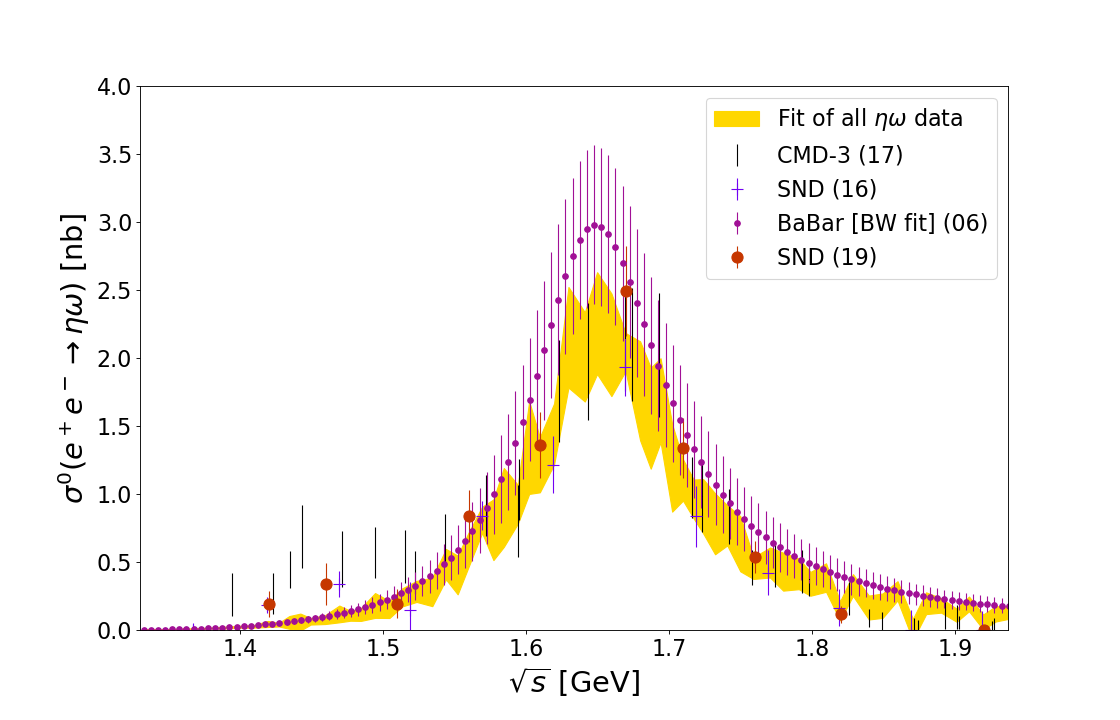}}\hfill
  \subfloat[$\sigma^{0}(e^+e^-\rightarrow\eta\phi)$]{%
    \includegraphics[width= 0.33\textwidth]{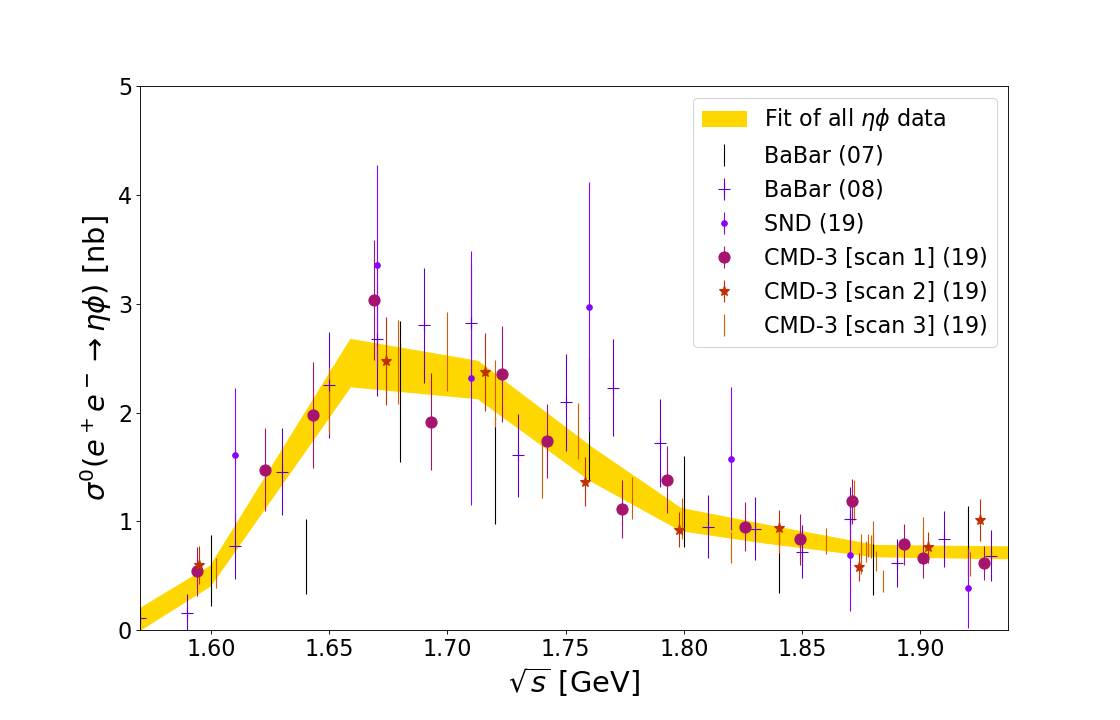}}\hfill
  \subfloat[$\sigma^{0}(e^+e^-\rightarrow\omega\eta\pi^0)$]{%
    \includegraphics[width= 0.33\textwidth]{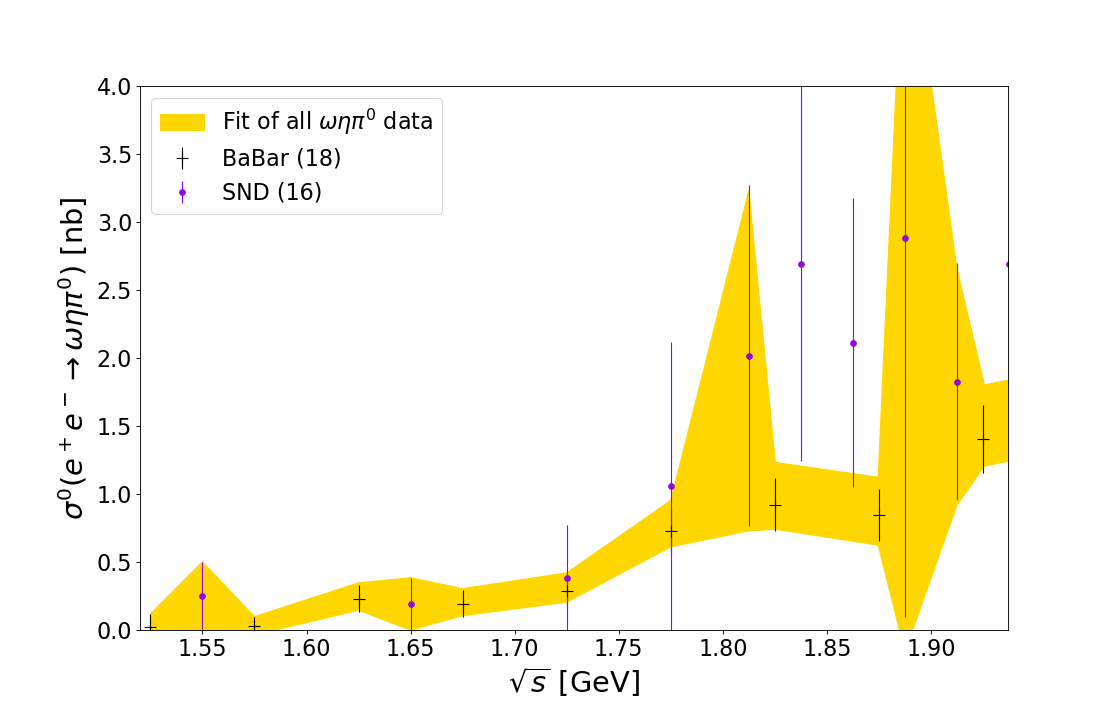}}\hfill
  \subfloat[$\sigma^{0}(e^+e^-\rightarrow2\pi^+2\pi^-\eta)$]{%
    \includegraphics[width= 0.33\textwidth]{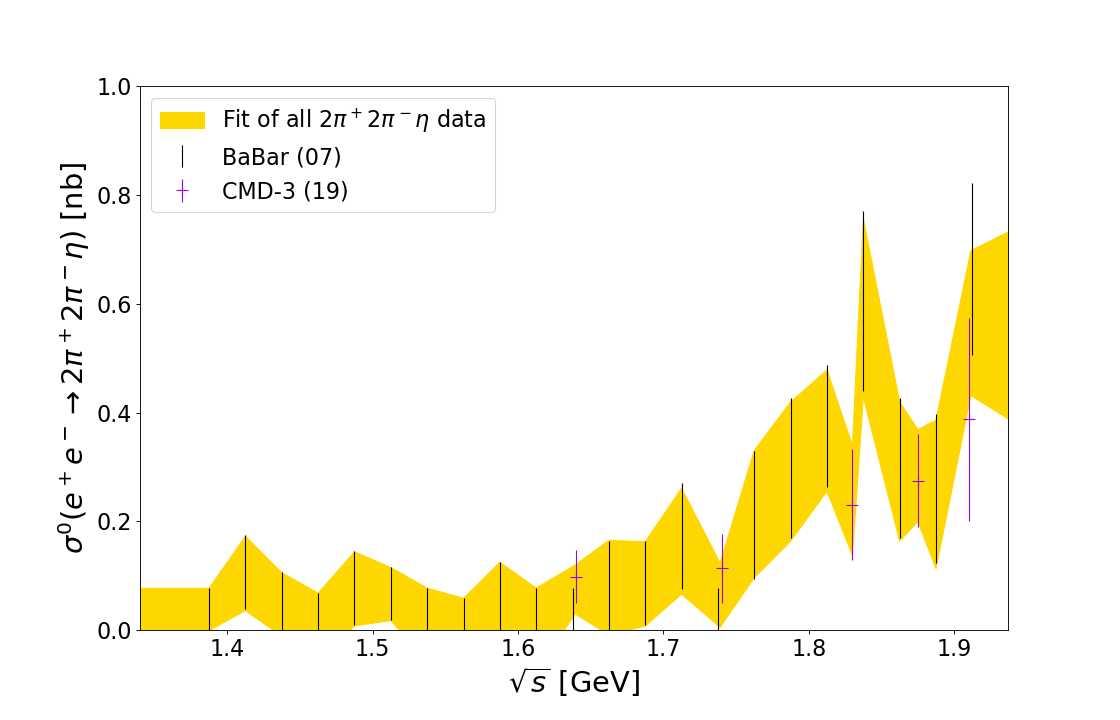}}\hfill
  \subfloat[$\sigma^{0}(e^+e^-\rightarrow2\pi^+2\pi^-\omega)$]{%
    \includegraphics[width= 0.33\textwidth]{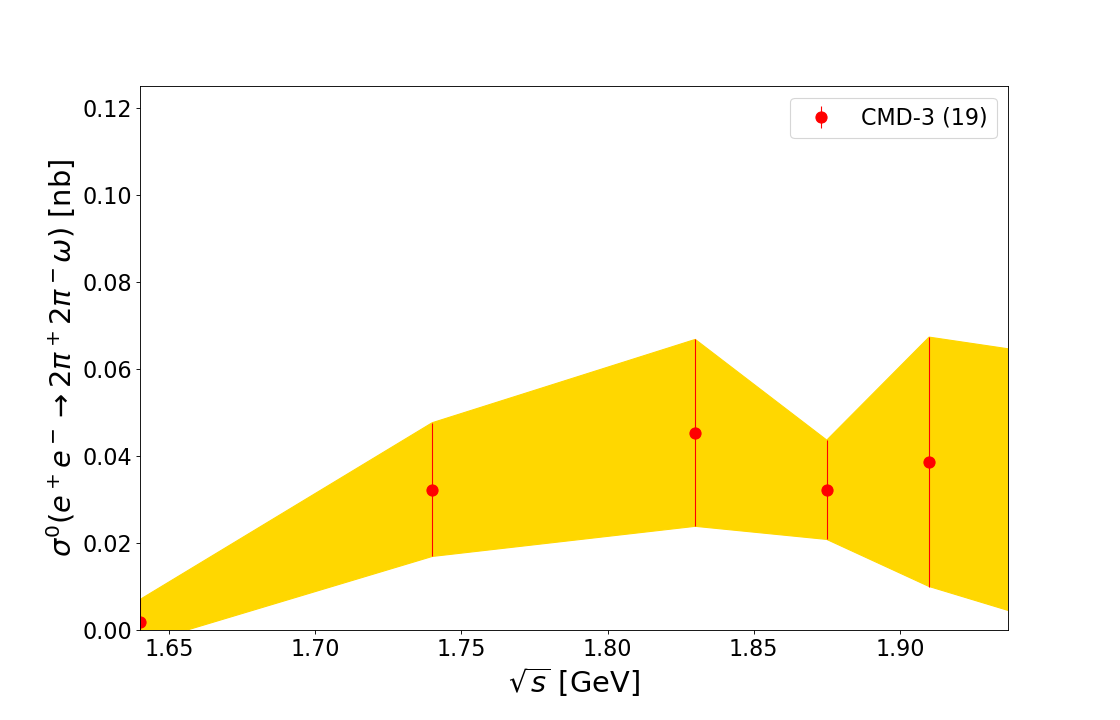}}\hfill
  \subfloat[$\sigma^{0}\big(e^+e^-\rightarrow(3\pi^+3\pi^-\pi^0)_{{\rm no}\, \eta\omega}\big)$]{%
    \includegraphics[width= 0.33\textwidth]{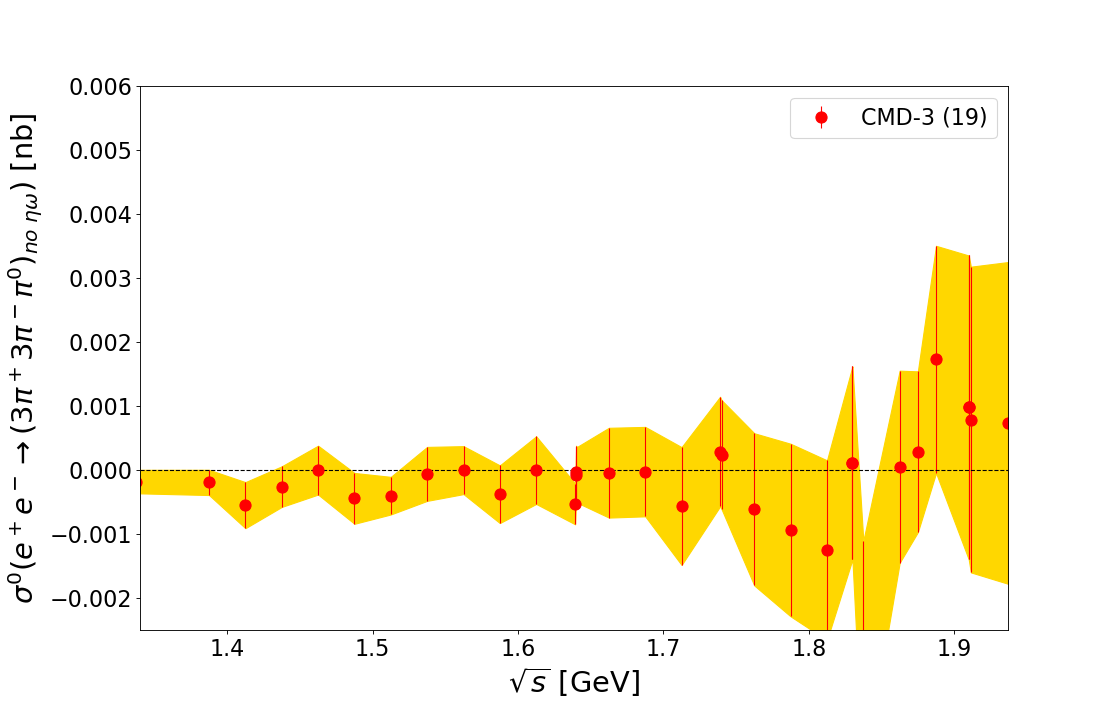}}\hfill
  \caption{The resulting cross sections of the updated, sub-leading hadronic channels contributing to the KNT19 data compilation.}\label{fig:otherContributions}
\end{figure}

\begin{figure}[!t]
\centering
  \subfloat[$\sigma^{0}\big(e^+e^-\rightarrow(\pi^+\pi^-3\pi^0)_{{\rm no}\, \eta}\big)$]{%
    \includegraphics[width= 0.33\textwidth]{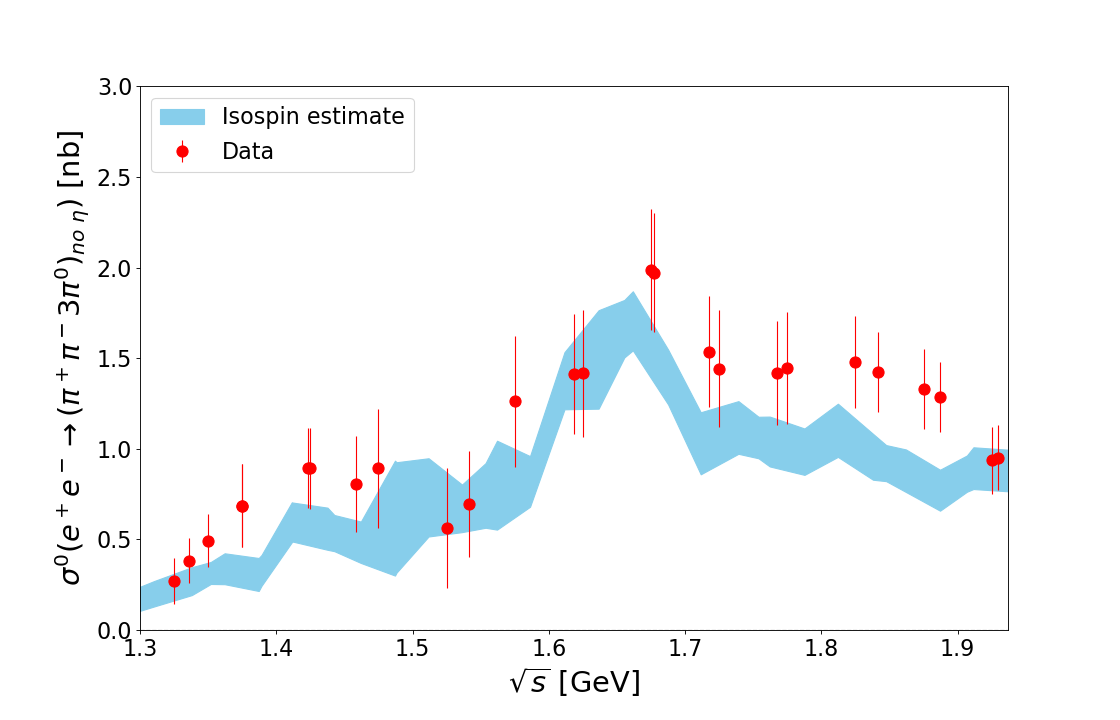}}\hfill
  \subfloat[$\sigma^{0}(e^+e^-\rightarrow \pi^+\pi^-2\pi^0\eta)$]{%
    \includegraphics[width= 0.33\textwidth]{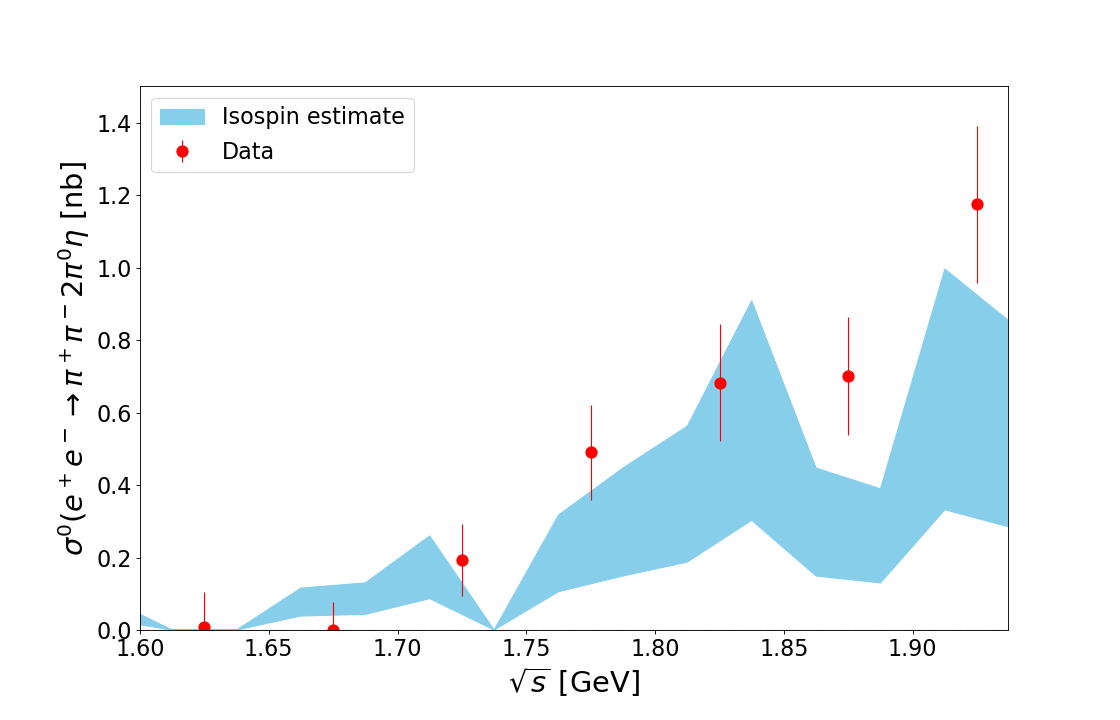}}\hfill
  \subfloat[$\sigma^{0}(e^+e^-\rightarrow \omega(\rightarrow{\rm npp})\pi\pi)$]{%
    \includegraphics[width= 0.33\textwidth]{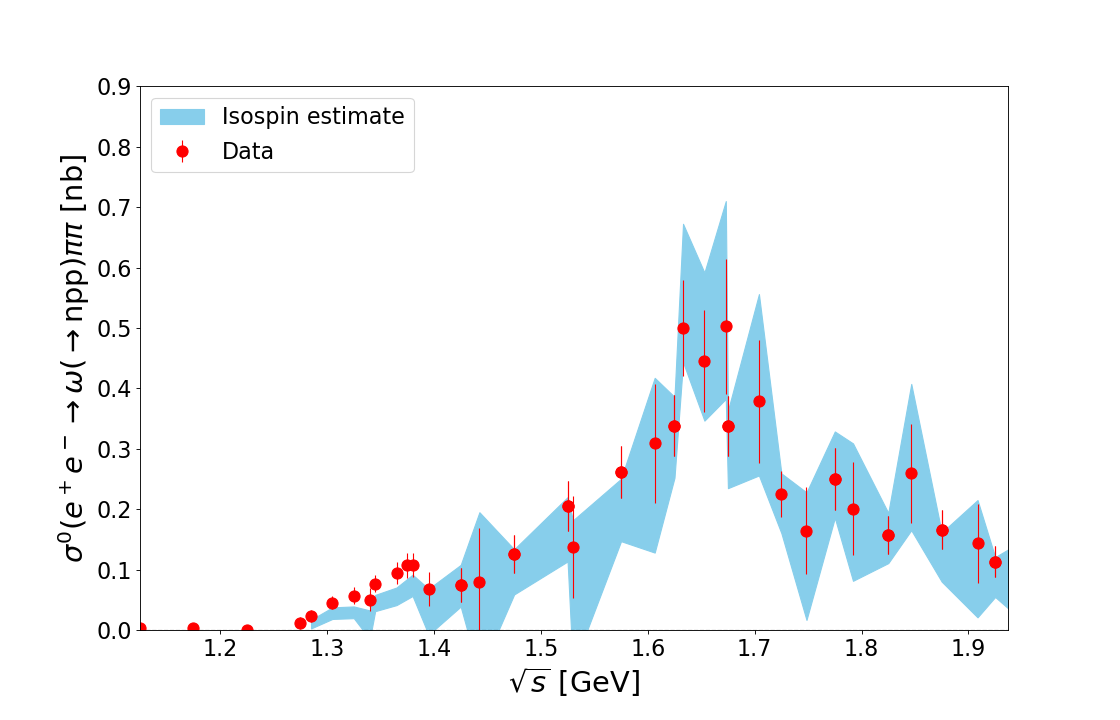}\label{fig:isospin_npp}}\hfill
  \caption{The resulting cross sections of those hadronic channels contributing to the KNT19 data compilation that were previously estimated via isospin relations. In Figure~\ref{fig:isospin}(c), the abbreviation `$\rightarrow$npp' represents the resonant decay to non-purely-pionic modes.}\label{fig:isospin}\vfill
\end{figure}

There have been a number of small data updates (see~\cite{Achasov:2018ujw,Lees:2018dnv,Gribanov:2019qgw,Achasov:2019duv,Ivanov:2019crp,CMD-3:2019ufp}) in other channels since~\cite{Keshavarzi:2018mgv}. The affected channels are all depicted in Figure~\ref{fig:otherContributions} and Figure~\ref{fig:isospin}. Notably, the $\pi^0\gamma$ channel now includes a new measurement from the SND experiment~\cite{Achasov:2018ujw}, which greatly extends the previous upper border of the channel from 1.35 GeV to 1.935 GeV in this work. The changes to $a_\mu^{\pi^0\gamma}$ are negligible, confirming that no higher energy contributions were missed previously in this hadronic mode. 

Two new channels are now included in the KNT19 data compilation. A
measurement of the $2\pi^+2\pi^-\omega$ channel by
CMD-3~\cite{CMD-3:2019ufp} provides a negligibly small addition to
$a_{\mu}^{\rm had, \, LO \, VP}$. This process, together with a
measurement of the $2\pi^+2\pi^-\eta$ mode, have provided the
production mechanisms to measure the seven-pion final state
$3\pi^+3\pi^-\pi^0$ in the same work~\cite{CMD-3:2019ufp}, which is
the first inclusion of a final state with more than six pions. After
removing the contributions from the $\eta$ and $\omega$ resonances to
avoid double-counting, the $3\pi^+3\pi^-\pi^0$ channel is
statistically consistent with zero below the upper energy boundary of
the sum of exclusive states used here, i.e.\ 1.937 GeV. Once again, it
is encouraging to ratify that no large contributions were missed from
these channels in the KNT18 data compilation. 

Lastly, it is important to mention that the three modes $\pi^+\pi^-3\pi^0$, $\pi^+\pi^-2\pi^0\eta$ and $\omega\pi^0\pi^0$ that were previously unmeasured have now been measured by BaBar~\cite{Lees:2018dnv}. These allow, for the first time, for their corresponding hadronic contributions to be estimated using experimental data instead of previously used isospin relations. All three channels are shown in Figure~\ref{fig:isospin}, where the agreement in each case between the data and the isospin prediction is good. The resulting integrated contributions to $a_{\mu}^{\rm had, \, LO \, VP}$ are all consistent with the theory estimates previously given in~\cite{Keshavarzi:2018mgv}.

\section{Results}\label{sec:Results}

 \begin{table}[htbp]
 \hspace{-0.8cm}
 \scalebox{0.8}{
 {\renewcommand{\arraystretch}{0.9}
 \begin{tabular}{|l|c|c|c|c|c|}
 \hline
 {\bf Channel} & $a_{e}^{\rm had, \, LO \, VP} \times 10^{14} $ & $a_{\mu}^{\rm had, \, LO \, VP} \times 10^{10}$ & $a_{\tau}^{\rm had, \, LO \, VP} \times 10^{8}$ & $\Delta\alpha_{\rm had}^{(5)}(M_Z^2) \times 10^{4}$ & $\Delta\nu_{\rm Mu}^{\rm had, \, VP}$ (Hz) \\
 \hline
 \multicolumn{6}{|c|}{Chiral perturbation theory (ChPT) threshold contributions} \\
 \hline
$\pi^0\gamma$                                                                                        & $      0.04\pm      0.00$ & $      0.12\pm      0.01$ & $      0.03\pm      0.00$ & $      0.00\pm      0.00$ & $      0.04\pm      0.00$ \\
$\pi^+\pi^-$                                                                                         & $      0.31\pm      0.01$ & $      0.87\pm      0.02$ & $      0.11\pm      0.00$ & $      0.01\pm      0.00$ & $      0.25\pm      0.01$ \\
$\pi^+\pi^-\pi^0$                                                                                    & $      0.00\pm      0.00$ & $      0.01\pm      0.00$ & $      0.00\pm      0.00$ & $      0.00\pm      0.00$ & $      0.00\pm      0.00$ \\
$\eta\gamma$                                                                                         & $      0.00\pm      0.00$ & $      0.00\pm      0.00$ & $      0.00\pm      0.00$ & $      0.00\pm      0.00$ & $      0.00\pm      0.00$ \\
 \hline
 \multicolumn{6}{|c|}{Exclusive channels ($\sqrt{s} \leq 1.937$ GeV)} \\
 \hline
$\pi^0\gamma$                                                                                        & $      1.19\pm      0.03$ & $      4.46\pm      0.10$ & $      1.75\pm      0.04$ & $      0.36\pm      0.01$ & $      1.45\pm      0.03$ \\
$\pi^+\pi^-$                                                                                         & $    138.59\pm      0.54$ \hphantom{,,\,} & $    503.46\pm      1.91$ \hphantom{,,\,} & $    172.84\pm      0.61$ \hphantom{,,\,} & $     34.29\pm      0.12$ \hphantom{,} & $    159.64\pm      0.60$ \hphantom{,,\,} \\
$\pi^+\pi^-\pi^0$                                                                                    & $     12.29\pm      0.25$ \hphantom{,} & $     46.73\pm      0.94$ \hphantom{,} & $     20.47\pm      0.39$ \hphantom{,} & $      4.69\pm      0.09$ & $     15.48\pm      0.31$ \hphantom{,} \\
$\pi^+\pi^-\pi^+\pi^-$                                                                               & $      3.67\pm      0.05$ & $     14.87\pm      0.20$ \hphantom{,} & $     11.50\pm      0.16$ \hphantom{,} & $      4.02\pm      0.05$ & $      5.58\pm      0.08$ \\
$\pi^+\pi^-\pi^0\pi^0$                                                                               & $      4.80\pm      0.19$ & $     19.39\pm      0.78$ \hphantom{,} & $     14.56\pm      0.58$ \hphantom{,} & $      5.00\pm      0.20$ & $      7.22\pm      0.29$ \\
$(2\pi^+2\pi^-\pi^0)_{{\rm no} \ \eta\omega}$                                                        & $      0.24\pm      0.02$ & $      0.98\pm      0.09$ & $      0.84\pm      0.08$ & $      0.32\pm      0.03$ & $      0.38\pm      0.03$ \\
$(\pi^+\pi^-3\pi^0)_{{\rm no} \ \eta}$                                                               & $      0.15\pm      0.03$ & $      0.62\pm      0.11$ & $      0.54\pm      0.10$ & $      0.21\pm      0.04$ & $      0.24\pm      0.04$ \\
$(3\pi^+3\pi^-)_{{\rm no} \ \omega}$                                                                 & $      0.06\pm      0.00$ & $      0.23\pm      0.01$ & $      0.21\pm      0.01$ & $      0.09\pm      0.01$ & $      0.09\pm      0.01$ \\
$(2\pi^+2\pi^-2\pi^0)_{{\rm no} \ \eta}$                                                             & $      0.33\pm      0.04$ & $      1.35\pm      0.17$ & $      1.24\pm      0.15$ & $      0.51\pm      0.06$ & $      0.53\pm      0.07$ \\
$(\pi^+\pi^-4\pi^0)_{{\rm no} \ \eta}$                                                               & $      0.05\pm      0.05$ & $      0.21\pm      0.21$ & $      0.19\pm      0.19$ & $      0.08\pm      0.08$ & $      0.08\pm      0.08$ \\
$(3\pi^+3\pi^-\pi^0)_{{\rm no} \ \eta\omega}$                                                        & $     0.00\pm      0.00$ & $     0.00\pm      0.01$ & $     0.00\pm      0.00$ & $     0.00\pm      0.00$ & $     0.00\pm      0.00$ \\
$K^+K^-$                                                                                             & $      5.86\pm      0.06$ & $     23.03\pm      0.22$ \hphantom{,} & $     12.82\pm      0.12$ \hphantom{,} & $      3.37\pm      0.03$ & $      8.01\pm      0.08$ \\
$K^0_SK^0_L$                                                                                         & $      3.33\pm      0.05$ & $     13.04\pm      0.19$ \hphantom{,} & $      7.00\pm      0.10$ & $      1.77\pm      0.03$ & $      4.51\pm      0.07$ \\
$KK\pi$                                                                                              & $      0.66\pm      0.03$ & $      2.71\pm      0.12$ & $      2.33\pm      0.10$ & $      0.89\pm      0.04$ & $      1.05\pm      0.05$ \\
$KK2\pi$                                                                                             & $      0.47\pm      0.02$ & $      1.93\pm      0.08$ & $      1.80\pm      0.07$ & $      0.75\pm      0.03$ & $      0.76\pm      0.03$ \\
$KK3\pi$                                                                                             & $      0.01\pm      0.00$ & $      0.04\pm      0.02$ & $      0.04\pm      0.02$ & $      0.02\pm      0.01$ & $      0.02\pm      0.01$ \\
$\eta\gamma$                                                                                         & $      0.18\pm      0.01$ & $      0.70\pm      0.02$ & $      0.35\pm      0.01$ & $      0.09\pm      0.00$ & $      0.24\pm      0.01$ \\
$\eta\pi^+\pi^-$                                                                                     & $      0.33\pm      0.01$ & $      1.34\pm      0.05$ & $      1.10\pm      0.04$ & $      0.41\pm      0.02$ & $      0.51\pm      0.02$ \\
$(\eta\pi^+\pi^-\pi^0)_{{\rm no} \ \omega}$                                                          & $      0.17\pm      0.02$ & $      0.71\pm      0.08$ & $      0.63\pm      0.07$ & $      0.25\pm      0.03$ & $      0.28\pm      0.03$ \\
$\eta2\pi^+2\pi^-$                                                                                   & $      0.02\pm      0.00$ & $      0.08\pm      0.01$ & $      0.07\pm      0.01$ & $      0.03\pm      0.00$ & $      0.03\pm      0.00$ \\
$\eta\pi^+\pi^-\pi^0\pi^0$                                                                           & $      0.03\pm      0.00$ & $      0.12\pm      0.02$ & $      0.11\pm      0.02$ & $      0.05\pm      0.01$ & $      0.05\pm      0.01$ \\
$\eta\omega$                                                                                         & $      0.07\pm      0.01$ & $      0.30\pm      0.02$ & $      0.26\pm      0.02$ & $      0.10\pm      0.01$ & $      0.11\pm      0.01$ \\
$\omega(\rightarrow\pi^0\gamma)\pi^0$                                                                & $      0.22\pm      0.00$ & $      0.88\pm      0.02$ & $      0.61\pm      0.01$ & $      0.19\pm      0.00$ & $      0.32\pm      0.01$ \\
$\omega(\rightarrow{\rm npp})2\pi$                                                                   & $      0.03\pm      0.00$ & $      0.13\pm      0.01$ & $      0.12\pm      0.01$ & $      0.04\pm      0.00$ & $      0.05\pm      0.01$ \\
$\omega(\rightarrow{\rm npp})3\pi$                                                                   & $      0.04\pm      0.01$ & $      0.17\pm      0.03$ & $      0.15\pm      0.03$ & $      0.06\pm      0.01$ & $      0.07\pm      0.01$ \\
$\omega2\pi^+2\pi^-$                                                                                 & $      0.00\pm      0.00$ & $      0.01\pm      0.00$ & $      0.01\pm      0.00$ & $      0.00\pm      0.00$ & $      0.00\pm      0.00$ \\
$\eta\phi$                                                                                           & $      0.10\pm      0.00$ & $      0.41\pm      0.02$ & $      0.37\pm      0.02$ & $      0.15\pm      0.01$ & $      0.16\pm      0.01$ \\
$\omega\eta\pi^0$                                                                                    & $      0.06\pm      0.01$ & $      0.24\pm      0.05$ & $      0.23\pm      0.05$ & $      0.10\pm      0.02$ & $      0.10\pm      0.02$ \\
$\omega(\rightarrow{\rm npp})KK$                                                                     & $      0.00\pm      0.00$ & $      0.00\pm      0.00$ & $      0.00\pm      0.00$ & $      0.00\pm      0.00$ & $      0.00\pm      0.00$ \\
$\eta(\rightarrow{\rm npp})KK_{{\rm no} \ \phi\rightarrow KK}$                                       & $      0.00\pm      0.00$ & $      0.01\pm      0.01$ & $      0.01\pm      0.01$ & $      0.01\pm      0.00$ & $      0.01\pm      0.01$ \\
$\phi\rightarrow{\rm unaccounted}$                                                                   & $      0.01\pm      0.01$ & $      0.04\pm      0.04$ & $      0.02\pm      0.02$ & $      0.01\pm      0.01$ & $      0.01\pm      0.01$ \\
$p\bar{p}$                                                                                           & $      0.01\pm      0.00$ & $      0.03\pm      0.00$ & $      0.03\pm      0.00$ & $      0.01\pm      0.00$ & $      0.01\pm      0.00$ \\
$n\bar{n}$                                                                                           & $      0.01\pm      0.00$ & $      0.03\pm      0.01$ & $      0.03\pm      0.01$ & $      0.01\pm      0.00$ & $      0.01\pm      0.00$ \\
 \hline
 \multicolumn{6}{|c|}{Other contributions ($\sqrt{s} > 1.937$ GeV)} \\
 \hline
Inclusive channel                                                                                    & $     10.38\pm      0.16$ \hphantom{\,} & $     43.55\pm      0.67$ \hphantom{\,} & $     63.49\pm      0.91$ \hphantom{\,} & $     82.78\pm      1.05$ \hphantom{\,} & $     19.82\pm      0.30$ \hphantom{\,} \\
$J/\psi$                                                                                             & $      1.49\pm      0.05$ & $      6.26\pm      0.19$ & $      8.91\pm      0.27$ & $      7.07\pm      0.22$ & $      2.81\pm      0.09$ \\
$\psi'$                                                                                              & $      0.37\pm      0.01$ & $      1.58\pm      0.04$ & $      2.50\pm      0.06$ & $      2.51\pm      0.06$ & $      0.74\pm      0.02$ \\
$\Upsilon(1S)$                                                                                       & $      0.01\pm      0.00$ & $      0.05\pm      0.00$ & $      0.12\pm      0.00$ & $      0.55\pm      0.02$ & $      0.03\pm      0.00$ \\
$\Upsilon(2S)$                                                                                       & $      0.00\pm      0.00$ & $      0.02\pm      0.00$ & $      0.05\pm      0.00$ & $      0.24\pm      0.01$ & $      0.01\pm      0.00$ \\
$\Upsilon(3S)$                                                                                       & $      0.00\pm      0.00$ & $      0.01\pm      0.00$ & $      0.03\pm      0.00$ & $      0.17\pm      0.01$ & $      0.01\pm      0.00$ \\
$\Upsilon(4S)$                                                                                       & $      0.00\pm      0.00$ & $      0.01\pm      0.00$ & $      0.02\pm      0.00$ & $      0.10\pm      0.01$ & $      0.00\pm      0.00$ \\
pQCD\ ($\sqrt{s}> 11.199$~GeV)                                                                & $      0.48\pm      0.00$ & $      2.07\pm      0.00$ & $      5.33\pm      0.00$ & $    124.79\pm      0.09$ \hphantom{,,} & $      1.34\pm      0.00$ \\
 \hline
 \hline
Total ($ <\infty$ GeV) &$    186.08\pm      0.66$ \hphantom{,,} & $    692.78\pm      2.42$ \hphantom{,,} & $    332.81\pm      1.39$ \hphantom{,,} & $    276.09\pm      1.12$ \hphantom{,,} & $    232.04\pm      0.82$ \hphantom{,,} \\
 \hline
 \end{tabular}
 }
 }
 \caption{Summary of the contributions to $a_{e}^{\rm had, \, LO \,
     VP}$, $a_{\mu}^{\rm had, \, LO \, VP}$, $a_{\tau}^{\rm had, \, LO
     \, VP}$, $\Delta\alpha_{\rm had}^{(5)}(M_Z^2)$ and
   $\Delta\nu_{\rm Mu}^{\rm had, \, VP}$ calculated in this
   analysis. The first column indicates the channel, the second, third
   and fourth columns give the contributions to $a_{e}^{\rm had, \, LO
     \, VP}$, $a_{\mu}^{\rm had, \, LO \, VP}$ and $a_{\tau}^{\rm had,
     \, LO \, VP}$, whereas the fifth and the last column list the
   contributions to $\Delta\alpha_{\rm had}^{(5)}(M_Z^2)$ and
   $\Delta\nu_{\rm Mu}^{\rm had, \, VP}$, respectively. The last row
   describes the total contribution obtained from the sum of the
   individual final states, with the uncertainties added in
   quadrature.}\label{tab:comparisonTable}
 \end{table}

Table~\ref{tab:comparisonTable} shows the contributions of the individual hadronic channels to $a_{e}^{\rm had, \, LO \, VP}$, $a_{\mu}^{\rm had, \, LO \, VP}$, $a_{\tau}^{\rm had, \, LO \, VP}$, $\Delta\alpha_{\rm had}^{(5)}(M_Z^2)$ and $\Delta\nu_{\rm Mu}^{\rm had, \, VP}$ calculated in this analysis. For $a_{l}^{\rm had, \, LO \, VP}$ $(l = e, \mu, \tau)$, the combined hadronic cross section data for each channel are integrated according to equation~\eqref{eq:amu}. To obtain $\Delta\alpha_{\rm had}^{(5)}(M_Z^2)$, the data are integrated using equation~\eqref{eq:delAlpha} given in Section~\ref{sec:alphaMz}. For $\Delta\nu_{\rm Mu}^{\rm had, \, VP}$, equation~\eqref{eq:HFShadVP} in Section~\ref{sec:hypMuon} is used. In the following section, the KNT19 results for $a_e$, $a_\mu$, $a_\tau$, $\alpha(M_{Z}^2)$ and $\Delta\nu_{\rm Mu}$ are presented separately. For each of the lepton $g-2$ results, the values for the LO and NLO hadronic VP contributions as calculated in this work are given, followed by corresponding updated estimates for the respective SM predictions and any necessary discussions. 

\subsection{The anomalous magnetic moment of the electron, $a_e$}

Integrating the updated KNT19 determination of the hadronic $R$-ratio described in Section~\ref{sec:DataUpdates} according to equation~\eqref{eq:amu} (with $l = e$) results in
\begin{align} \label{aeLOHVP_KNT19}
a_{e}^{\rm had, \, LO \, VP} & = (186.08  \pm 0.34_{\rm stat} \pm 0.53_{\rm sys} \pm 0.05_{\rm vp} \pm 0.18_{\rm fsr}) \times 10^{-14}  \nonumber
\\
\
& =  (186.08 \pm 0.66_{\rm tot}) \times 10^{-14} \ .
\end{align}
The contributions from the individual hadronic channels contributing to $a_{e}^{\rm had, \, LO \, VP}$ are listed in Table~\ref{tab:comparisonTable}. With the same data input, the NLO contributions to $a_{e}^{\rm had,\, VP}$ are determined here to be
\begin{align}\label{aeNLOHVP_KNT19}
a_{e}^{\rm had,\, NLO\, VP} & = (-22.28  \pm 0.04_{\rm stat} \pm 0.06_{\rm sys} \pm 0.01_{\rm vp} \pm 0.02_{\rm fsr}) \times 10^{-14}  \nonumber
\\
\
& =  (-22.28 \pm 0.08_{\rm tot}) \times 10^{-14} \ .
\end{align}
The NT12 analysis ~\cite{Nomura:2012sb} found $a_{e}^{\rm had, \, LO
  \, VP}({\rm NT12}) = (186.6 \pm 1.1) \times 10^{-14}$ and
$a_{e}^{\rm had, \, NLO \, VP}({\rm NT12}) = (-22.34 \pm 0.14) \times
10^{-14}$. Comparing the results in this analysis with those from
NT12, the mean values have decreased by a substantial fraction of the
previously quoted uncertainties (although well within them) and the
uncertainties themselves have reduced by $> 40\%$. This is in line with the changes noted in the KNT18 determination of $a_\mu$~\cite{Keshavarzi:2018mgv}, which observed similar changes largely due to reductions in the mean value and uncertainty of the dominant $\pi^+\pi^-$ channel. 

As the NNLO hadronic VP contributions are not calculated in this work, 
the result \allowbreak $a_{e}^{\rm had, \, NNLO \, VP} = (2.80 \pm 0.01) \times
10^{-14}$ from~\cite{Kurz:2014wya} is adopted which utilises the 
HLMNT11~\cite{Hagiwara:2011af} data compilation for the hadronic $R$-ratio.\footnote{During the KNT18 analysis, the authors of~\cite{Kurz:2014wya} 
kindly repeated their analysis with the KNT18 data compilation and found 
negligible changes with respect to their published result.} For the hadronic LbL contributions,
the value $a_{e}^{\rm had, \, LbL} = (3.7 \pm 0.5) \times 10^{-14}$
from~\cite{Jegerlehner:2017zsb} is used. With these, the full hadronic
contributions to the electron $g-2$ are estimated to be 
\begin{table}[!t]
\hspace{-0.4cm}
\scalebox{0.85}{
{\renewcommand{\arraystretch}{1.1}
\begin{tabular}{|l|r r r|}
\hline 
{\bf SM contribution} & \multicolumn{1}{c}{$a_{e} (\alpha_{\rm Rb}) \times 10^{12}$} & &  \multicolumn{1}{c|}{$a_{e} (\alpha_{\rm Cs}) \times 10^{12}$}  \\
\hline 
QED & $ 1159652180.309 \pm 0.720 $~\cite{Aoyama:2017uqe} & & $ 1159652179.887 \pm 0.230 $~\cite{Parker:2018vye} \\
EW &  & $ 0.031 \pm 0.000$~\cite{Aoyama:2017uqe} &  \\
had LO VP &   & $1.861 \pm 0.007$~\hphantom{[12]} &   \\
had NLO VP &  & $  -0.223 \pm 0.001 $~\hphantom{[12]} &   \\
had NNLO VP &  & $  0.028 \pm 0.000 $~\cite{Aoyama:2017uqe} &  \\
had LbL &  & $  0.037 \pm 0.005  $~\cite{Aoyama:2017uqe} &  \\
\hline
Theory total & $ 1159652182.042 \pm 0.720 $ & & $1159652181.620 \pm 0.230$ \\ 
Experiment &  & $  1159652180.730 \pm 0.280 $~\cite{Hanneke:2008tm} &  \\ 
\hline 
$\Delta a_{e}$ & $  -1.312 \pm 0.773  \ (1.7\sigma) $ & & $ -0.890 \pm 0.362  \ (2.5\sigma)$   \\ 
\hline 
\end{tabular} 
}
}\caption{Summary of the contributions to $a_{e}^{\rm SM}$. The values of $a_e^{\rm QED}$ from $\alpha_{\rm Rb}$ (left) and $\alpha_{\rm Cs}$ (right) and their resulting values for $a_e^{\rm SM}$ and $\Delta a_{e}$ are listed individually for comparison. All results are given as $a_{e}^{\rm SM} \times 10^{12}$.} \label{tab:aeSM-KNT19} 
\end{table}
\begin{figure}[!t] 
  \centering
    \includegraphics[width=0.82\textwidth]{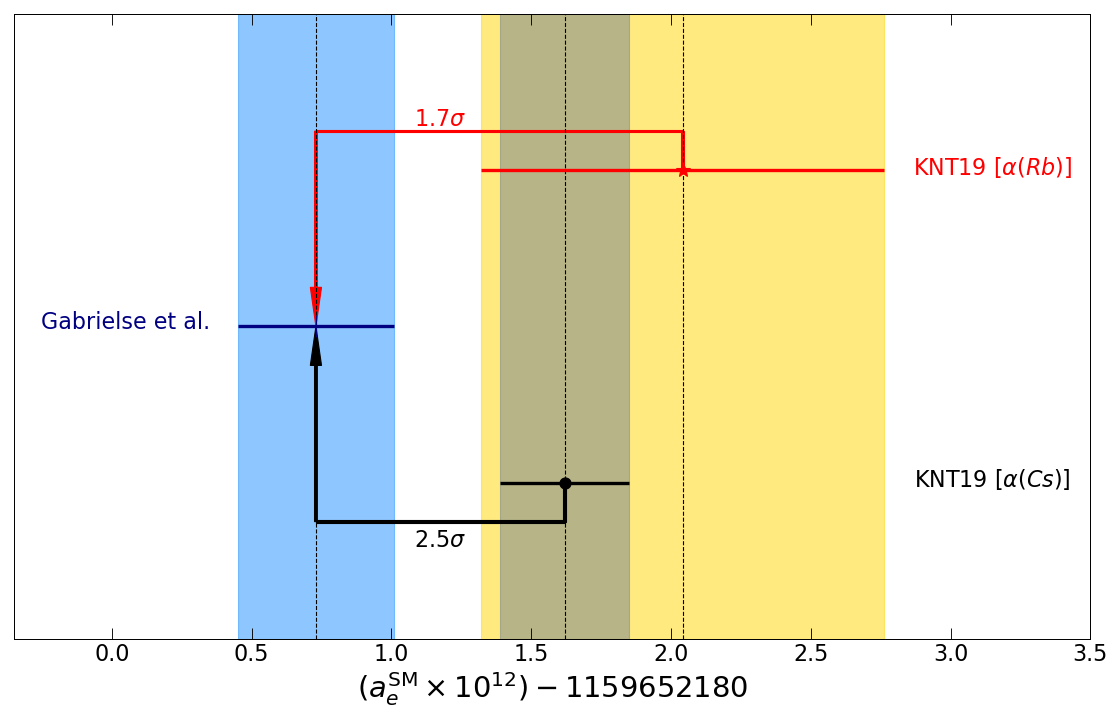}
     \caption{\small A comparison of the evaluations of $a_{e}^{\rm
         SM}$ as determined in this work with the experimental
       measurement by Gabrielse {\it et al.}~\cite{Hanneke:2008tm},
       the uncertainty of which is given by the light blue band. The
       red marker and yellow band denote the determination of
       $a_{e}^{\rm SM}$ using $\alpha_{\rm Rb}$, whilst the black
       marker and grey band denote the determination of $a_{e}^{\rm
         SM}$ using $\alpha_{\rm Cs}$ (for the values see equations~\eqref{aeSM} 
       or Table~\ref{tab:aeSM-KNT19}).}     \label{aeSMCompare} 
\end{figure} 
\beq\label{aehad_KNT19}
a_{e}^{\rm had} =  (170.30 \pm 0.77_{\rm tot}) \times 10^{-14} \,,
\eeq
where, due to the complete correlations from the same input $R$-ratio,
the errors of the hadronic VP contributions have been added linearly.
Compared to $a_{e}^{\rm had}({\rm NT12}) =  (167.8 \pm 1.4) \times
10^{-14}$ in~\cite{Nomura:2012sb}, the mean value found in this work
is outside the quoted error given in~\cite{Nomura:2012sb}. However, it
should be noted that no determination of the NNLO hadronic VP contributions
was available for~\cite{Nomura:2012sb}, whereas in this work the
addition of $a_{e}^{\rm had, \, NNLO \, VP} = (2.80 \pm 0.01) \times
10^{-14}$ constitutes, similar to the case of the muon, a significant
additional correction.

The EW contributions, $a_{e}^{\rm EW} =  (3.053 \pm 0.023) \times
10^{-14}$, are also taken from~\cite{Jegerlehner:2017zsb}. For the QED
contributions, there are now two options depending on the choice for
the value of $\alpha$.\footnote{For the contributions from all the
  other sectors of the SM, the changes from the choice of $\alpha$ are
  negligible.} As described in Section~\ref{sec:Intro}, the use of the
measurement of $\alpha$ from Rb atomic
interferometry~\cite{Bouchendira:2010es} or Cs atomic
interferometry~\cite{Parker:2018vye} leads to an interesting
comparison with $a_e^{\rm exp}$. For each case, the values of
$a_e^{\rm QED}$ are 
\begin{align}\label{aeQED}
a_{e}^{\rm QED}(\alpha_{\rm Rb}) & = (115965218030.9 \pm 72.0) \times 10^{-14}\text{~\cite{Aoyama:2017uqe}} \,, \nonumber
\\
\
a_{e}^{\rm QED}(\alpha_{\rm Cs}) & =  (115965217988.7 \pm 23.0) \times 10^{-14}\text{~\cite{Parker:2018vye}} \ .
\end{align}
Using these and the contributions from the EW and hadronic sectors, the SM predictions for $a_e$ are found here to be
\begin{align}\label{aeSM}
a_{e}^{\rm SM}(\alpha_{\rm Rb}) & = (1159652182.042 \pm 0.72) \times 10^{-12} \,, \nonumber
\\
\
a_{e}^{\rm SM}(\alpha_{\rm Cs}) & =  (1159652181.620 \pm 0.23) \times 10^{-12} \ .
\end{align}
The comparison of these results with the experimental measurement of $a_e$~\cite{Hanneke:2008tm} is given in Table~\ref{tab:aeSM-KNT19} and shown in Figure~\ref{aeSMCompare}. The values of the deviation between theory and experiment of $\Delta a_{e}(\alpha_{\rm Rb}) = (-1.31 \pm 0.77) \times 10^{-12} \ (1.7\sigma)$ and $\Delta a_{e}(\alpha_{\rm Cs}) = (-0.89 \pm 0.36) \times 10^{-12} \ (2.5\sigma)$ confirm the findings in~\cite{Aoyama:2017uqe} and~\cite{Parker:2018vye}, respectively.

\subsection{The anomalous magnetic moment of the muon, $a_\mu$}

\begin{figure}[!t] 
  \centering
    \includegraphics[width=0.82\textwidth]{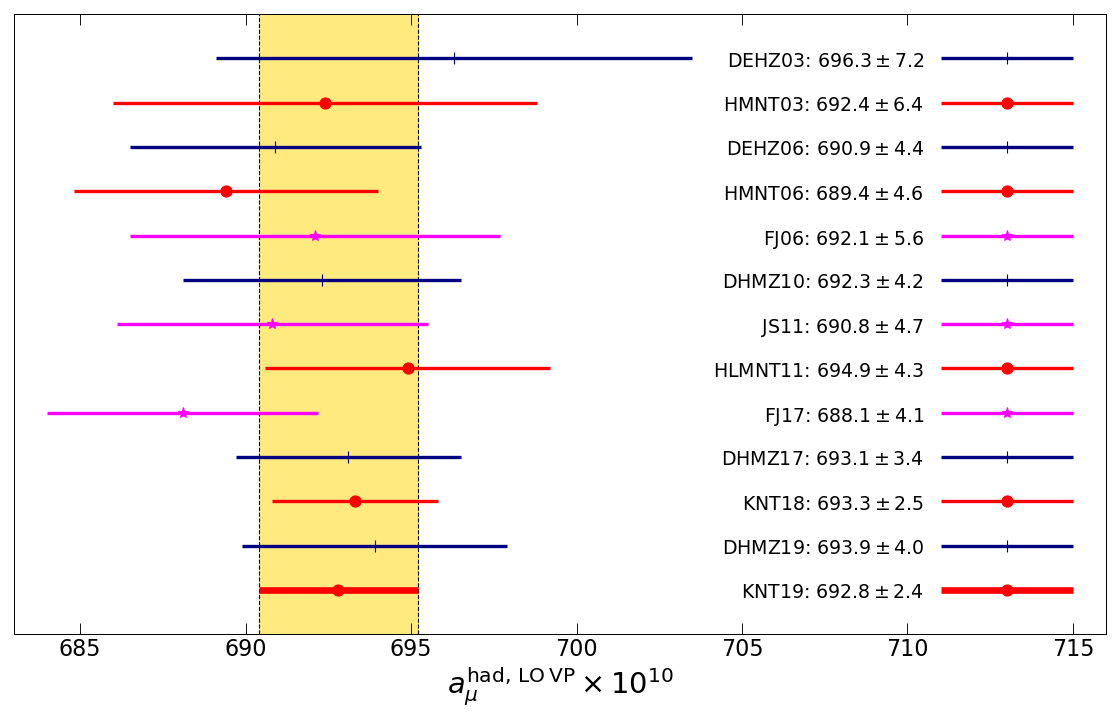}
     \caption{\small Comparison of recent and previous evaluations of
       $a_{\mu}^{\rm had, \, LO \, VP}$ determined from
       $e^+e^-\rightarrow {\rm hadrons}$ cross section data. The
       analyses listed in chronological order are:
       DEHZ03~\cite{Davier:2003pw}, HMNT03~\cite{Hagiwara:2003da},
       DEHZ06~\cite{Davier:2007ua}, HMNT06~\cite{Hagiwara:2006jt},
       FJ06~\cite{Jegerlehner:2006ju}, DHMZ10~\cite{Davier:2010nc},
       JS11~\cite{Jegerlehner:2011ti}, HLMNT11~\cite{Hagiwara:2011af},
       FJ17~\cite{Jegerlehner:2017lbd}, DHMZ17~\cite{Davier:2017zfy},
       KNT18~\cite{Keshavarzi:2018mgv} and
       DHMZ19~\cite{Davier:2019can}. The prediction from this work is
       listed as KNT19 and defines the (yellow) uncertainty band shown
       for the comparison with the other analyses.}     \label{amuHVPCompare}
\end{figure}

For the hadronic VP contribution to $a_\mu$, at LO this analysis finds
\begin{align} \label{amuLOHVP_KNT19}
a_{\mu}^{\rm had, \, LO \, VP} & = (692.78  \pm 1.21_{\rm stat} \pm 1.97_{\rm sys} \pm 0.21_{\rm vp} \pm 0.70_{\rm fsr}) \times 10^{-10}  \nonumber
\\
\
& =  (692.78 \pm 2.42_{\rm tot}) \times 10^{-10} \,, 
\end{align}
and the NLO contributions are determined here to be
\begin{align}\label{amuNLOHVP_KNT19}
a_{\mu}^{\rm had,\, NLO\, VP} & = (-9.83  \pm 0.01_{\rm stat} \pm 0.03_{\rm sys} \pm 0.01_{\rm vp} \pm 0.02_{\rm fsr}) \times 10^{-10}  \nonumber
\\
\
& =  (-9.83 \pm 0.04_{\rm tot}) \times 10^{-10} \ .
\end{align}
These results are consistent with the KNT18 analysis. At LO,
the integral over the hadronic $R$-ratio determined
in~\cite{Keshavarzi:2018mgv} resulted in \allowbreak $a_{\mu}^{\rm
  had, \, LO \, VP}({\rm KNT18}) = (693.26 \pm 2.46) \times 10^{-10}$. Comparing this
with equation~\eqref{amuLOHVP_KNT19}, the reduction in the mean value
comes entirely from the updated treatment of the $\omega$ resonance in
the $\pi^+\pi^-\pi^0$ channel described in Section~\ref{sec:3pi}. This
change counteracts the small increase in the mean value from the
$\pi^+\pi^-$ channel due to the inclusion of the CLEO-c
data~\cite{Xiao:2017dqv} detailed in Section~\ref{chap:pipi}, as well
as the very small increase due to the newly included channels reported
in Section~\ref{sec:OtherChannels}. The marginal decrease in the
overall uncertainty is also due to the inclusion of the CLEO-c
data~\cite{Xiao:2017dqv}, which as explained previously has caused a
small decrease in the local $\chi^2$ error inflation of the dominant
two-pion contribution. A comparison of this result with similar
evaluations of $a_{\mu}^{\rm had, \, LO \, VP}$ determined from
$e^+e^-\rightarrow {\rm hadrons}$ cross section data is shown in
Figure~\ref{amuHVPCompare}. It is important to note that there is
clear stability and overall agreement between the different
analyses/groups over the consecutive years, despite contrasting
choices the different groups have made concerning how to treat the
hadronic cross section data, where to use perturbative QCD (pQCD) instead of
data and the application of other possible theoretical constraints.\footnote{
The most recent update from DHMZ19 has a larger uncertainty compared to that of DHMZ17, since DHMZ19 have included an additional error to account for the difference they obtain for $a_\mu^{\pi^+\pi^-}$ when discarding either the KLOE or the BaBar data. As the KNT $\pi^+\pi^-$ data combination benefits from stronger constraints imposed by the correlated uncertainties, the difference observed in $a_\mu^{\pi^+\pi^-}$ when discarding the data from either experiment is less severe. Therefore, and remembering also that data tensions are quantitatively accounted for in the resulting cross section by the local $\chi^2$ error inflation, no additional uncertainty for $a_\mu$ is applied in this analysis.}

Combining the results (\ref{amuLOHVP_KNT19}) and (\ref{amuNLOHVP_KNT19}) with the NNLO corrections, $a_{\mu}^{\rm had,\, NNLO\, VP} = (1.24 \pm 0.01) \times 10^{-10}$~\cite{Kurz:2014wya}, the total hadronic VP contribution to $a_\mu$ is estimated to be
\begin{align}\label{amuHVP_KNT19}
a_{\mu}^{\rm had,\, VP} & =  (684.19 \pm 2.38_{\rm tot}) \times 10^{-10} \,,
\end{align}
where, as in the case of the electron, the errors have been
added linearly due to the full correlation between the 
$R$-ratio input for the three contributions.
When considering the SM prediction, in the case of the muon ($l =
\mu$), the other contributions in equation~\eqref{alSMeq} require
reconsideration. 
In contrast to the case of the electron, the muon is, at the current
level of accuracy, not sensitive to the choice of either $\alpha({\rm
  Rb})$ or $\alpha({\rm Cs})$, or the updated five-loop QED contributions
from~\cite{Aoyama:2017uqe}. Hence the value of the QED contributions, 
to the accuracy needed and quoted here, is unchanged at
$a_{\mu}^{\rm QED} = (11658471.90 \pm 0.01) \times 10^{-10}$~\cite{Aoyama:2012wk,Aoyama:2017uqe}.
For the EW contributions, the value chosen here is also the same as
in~\cite{Keshavarzi:2018mgv}. However, it should be noted that an
independent numerical evaluation of the two-loop EW contributions was
recently performed~\cite{Ishikawa:2018rlv}, resulting in an estimate
of the total EW contributions of $a_{\mu}^{\rm EW} = ( 15.29 \pm 0.10
) \times 10^{-10}$. This is consistent with the previously chosen
value of $a_{\mu}^{\rm EW} = ( 15.36 \pm 0.10 ) \times
10^{-10}$~\cite{Gnendiger:2013pva} and therefore no adjustment is
made for this analysis.

For the hadronic LbL sector, in~\cite{Keshavarzi:2018mgv} the
commonly quoted `Glasgow consensus' estimate of $a_{\mu}^{\rm had, \,
  LbL}(\text{`Glasgow consensus'})  =  (10.5  \pm 2.6) \times
10^{-10}$~\cite{Prades:2009tw} was used, adjusted for a re-evaluation of the
contribution to $a_{\mu}^{\rm had, LbL}$ due to axial
exchanges~\cite{Jegerlehner:2015stw,Pauk:2014rta,Nyffeler:2016gnb}. This
led to $a_{\mu}^{\rm had, \, LbL} =  (9.8  \pm 2.6) \times
10^{-10}$~\cite{Nyffeler:2016gnb} being adopted for the KNT18
analysis. Since that time, the progress in determining
$a_{\mu}^{\rm had, LbL}$ using dispersive approaches (where dispersion
relations are formulated that allow for the determination of the
hadronic LbL contributions from experimental data) has been
significant.\footnote{This advancement
  has been largely influenced by the efforts of the Muon $g-2$ Theory
  Initiative~\cite{TGm2} and the commendable work and successes of the
  groups within it, which have formed the basis for the following
  choices for $a_{\mu}^{\rm had, \, LbL}$ made in this work.}
These determinations are of particular interest for this analysis, as the
fundamental approach to this work (and the works preceding
it~\cite{Keshavarzi:2018mgv,Hagiwara:2003da,Hagiwara:2006jt,Hagiwara:2011af})
is that any estimates given be as model-independent and/or as
data-driven as possible. With the contributions to the `Glasgow
consensus' estimate having been solely determined through
model-dependent approaches, moving towards data-based evaluations of
the hadronic LbL contributions is consistent with the general
methodology of this undertaking. 

Those hadronic LbL contributions that have been determined by
dispersive techniques are the pseudoscalar poles
($\pi^0,\eta,\eta'$)~\cite{Hoferichter:2018dmo,Masjuan:2017tvw,Hoferichter:2018kwz}, the
pion/kaon-box contributions~\cite{Colangelo:2017fiz,TGm2} and the
$S$-wave $\pi\pi$ rescattering contributions~\cite{Colangelo:2017fiz,Colangelo:2017qdm}.
In addition, a new analysis of (longitudinal) short distance constraints has very recently become available~\cite{Colangelo:2019uex, Colangelo:2019lpu}, complementing the dispersive determination of the pseudoscalar contributions.
The values for these contributions and their counterparts from the `Glasgow consensus'
estimate are shown in Table~\ref{tab:hadLbLCompare}, where the estimate of the pseudoscalar contributions of the `Glasgow consensus' already contains short distance contributions.
With the aim to
strive for a more model-independent approach, the value for
$a_{\mu}^{\rm had, \, LbL}$ in this work is taken as the sum of the
contributions determined via dispersive approaches, the new estimates of short distance and charm quark corrections, plus the sum of the
contributions from scalars, tensors and axial-vectors remaining from the original `Glasgow consensus' 
estimate.\footnote{Note that the adjustments of the axial
 contributions mentioned above and adopted
 in~\cite{Keshavarzi:2018mgv}, have recently been found not
 justified, see~\cite{MHoferichterSeattle2019}, hence the
 estimate for the axial contributions from the original `Glasgow
 consensus' is used here.}
This results in a value for the total hadronic LbL contribution of $a_{\mu}^{\rm had, \, LbL} = (9.34 \pm 2.92) \times 10^{-10}$, where the errors from the individual contributions have been summed linearly. This provides a conservative estimate of the overall uncertainty and also accounts for currently unavailable transverse short distance constraints, which are estimated to be sub-leading.

\begin{table}[!t]
\centering
\addtolength{\tabcolsep}{0.1cm}
{\renewcommand{\arraystretch}{1.2}
\begin{tabular}{| c | c  c |}
\hline
Contribution & `Glasgow consensus'~\cite{Prades:2009tw} & Dispersive evaluations \\
\hline
$\pi^0,\eta,\eta'$-poles & $114\pm13$ \hphantom{1} & \hphantom{$-$\,~\cite{Hoferichter:2018dmo,Masjuan:2017tvw}}$93.8\pm4.0$~\cite{Hoferichter:2018dmo,Masjuan:2017tvw,Hoferichter:2018kwz} \\
$\pi/K$-box & $-19\pm19$ \hphantom{$-$} &  \hphantom{~\cite{Colangelo:2017fiz,TGm2}}$-16.4\pm0.2$~\cite{Colangelo:2017fiz,TGm2} \\
$S$-wave $\pi\pi$ rescattering & - &  \hphantom{~\cite{Colangelo:2017fiz,Colangelo:2017qdm},,}$-8\pm1$~\cite{Colangelo:2017fiz,Colangelo:2017qdm} \\
Short-distance contributions & [Part of $\pi^0,\eta,\eta'$-poles] & \hphantom{$-$~\cite{Hoferichter:2018dmo,Masjuan:2017tvw}.0} $13 \pm 6$~\cite{Colangelo:2019uex, Colangelo:2019lpu}  \hphantom{.0} \\
Charm contributions & 2.3 &  \hphantom{$-$~\cite{Hoferichter:2018dmo,Masjuan:2017tvw}..00}$3 \pm 1$~\cite{Colangelo:2019uex, Colangelo:2019lpu} \hphantom{.0}  \\
Scalars \& Tensors & \multicolumn{2}{c|}{$-7\pm7$}  \\
Axial-vectors & \multicolumn{2}{c|}{$15\pm10$} \\

\hline
Total & $105 \pm 26$ \hphantom{1} & \hphantom{$-$} $93.4 \pm 29.2$ \\
\hline
\end{tabular}
}\caption{Comparison of the contributions to $a_{\mu}^{\rm had, \, LbL}$ from the `Glasgow consensus' estimate and from recent evaluations mainly based on dispersive approaches. The single column results from the scalars, tensors and axial-vectors originate from the `Glasgow consensus' estimate. The total uncertainty for the value including the dispersive evaluations is determined via the conservative linear sum of the errors of the individual contributions. All results are given as $a_{\mu}^{\rm had, \, LbL}\times10^{11}$.}\label{tab:hadLbLCompare} 
\end{table}

\begin{figure}[!t] 
  \centering
    \includegraphics[width=0.82\textwidth]{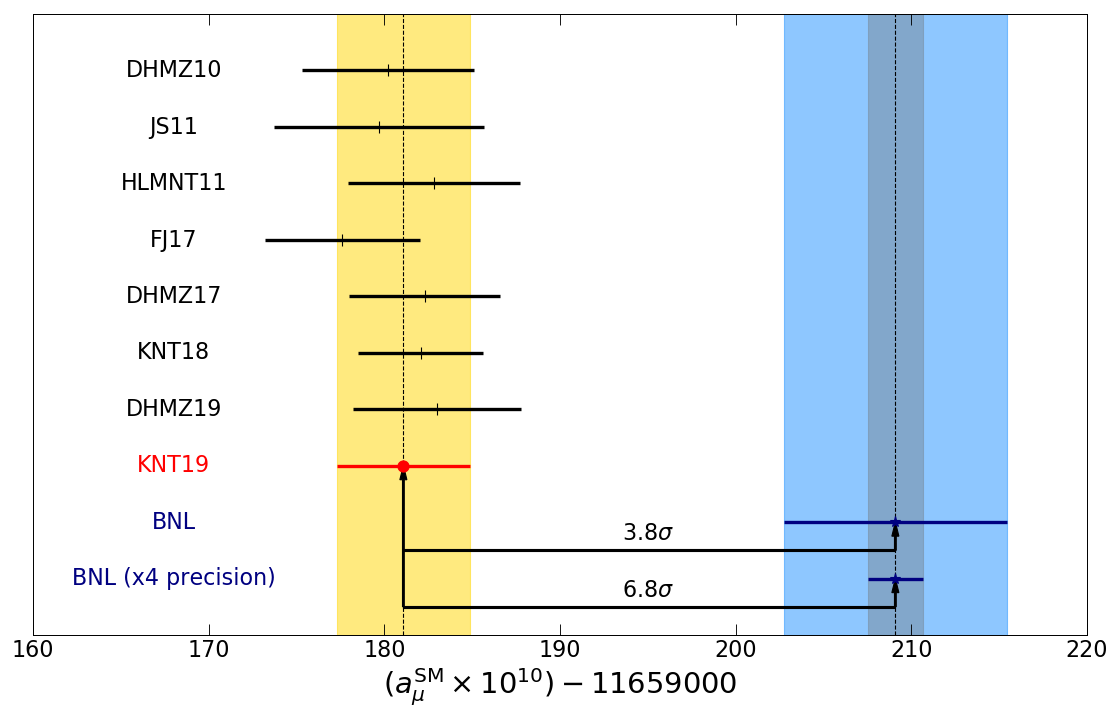}
     \caption{\small A comparison of recent and previous evaluations of $a_{\mu}^{\rm SM}$. The analyses listed in chronological order are: DHMZ10~\cite{Davier:2010nc}, JS11~\cite{Jegerlehner:2011ti}, HLMNT11~\cite{Hagiwara:2011af}, FJ17~\cite{Jegerlehner:2017lbd} and DHMZ17~\cite{Davier:2017zfy}, KNT18~\cite{Keshavarzi:2018mgv} and DHMZ19~\cite{Davier:2019can}. The prediction from this work is listed as KNT19, which defines the uncertainty band that other analyses are compared to. The current uncertainty on the experimental measurement~\cite{Bennett:2002jb,Bennett:2004pv,Bennett:2006fi,PDG2018} is given by the light blue band. The light grey band represents the hypothetical situation of the new experimental measurement at Fermilab yielding the same mean value for $a_{\mu}^{\rm exp}$ as the BNL measurement, but achieving the projected four-fold improvement in its uncertainty~\cite{Grange:2015fou}.}     \label{amuCompare}
\end{figure}

The values for the contributions from all the individual sectors of the SM chosen in this analysis are summarised in Table~\ref{tab:amuSM-KNT19}. 
\begin{table}[!t]
\centering
\scalebox{1.0}{
{\renewcommand{\arraystretch}{1.1}
\begin{tabular}{|l|c|}
\hline 
{\bf SM contribution} &  $a_{\mu} \times 10^{10}$  \\
\hline 
QED & $ 11658471.90 \pm 0.01 $~\cite{Aoyama:2017uqe} \\
EW & $ \hphantom{116584}15.36 \pm 0.10 $~\cite{Gnendiger:2013pva} \\
had LO VP & $ \hphantom{11658} 692.78 \pm 2.42 $~\hphantom{[12]}   \\
had NLO VP & $ \hphantom{116584} -9.83 \pm 0.04 \hphantom{-} $~\hphantom{[12]}   \\
had NNLO VP & $ \hphantom{1165847} 1.24 \pm 0.01 $~\cite{Kurz:2014wya}  \\
had LO LbL & $ \hphantom{1165847}   9.34 \pm 2.92   $~\hphantom{[12]}   \\
had NLO LbL & $ \hphantom{1165847} 0.30 \pm 0.20 $~\cite{Colangelo:2014qya}  \\
\hline
Theory total & $ 11659181.08 \pm 3.78 $~\hphantom{[12]}  \\ 
Experiment & $ 11659209.10 \pm 6.33 $~\cite{Bennett:2006fi}\hphantom{,,}  \\ 
\hline 
$\Delta a_{\mu}$ & $ \hphantom{116584 \ (4.13\sigma)} 28.02 \pm 7.37  \ (3.80\sigma) $~\hphantom{[12]}  \\ 
\hline 
\end{tabular} 
}
}\caption{Summary of the contributions to $a_{\mu}^{\rm SM}$.}\label{tab:amuSM-KNT19} 
\end{table}
Summing these contributions together results in an updated SM prediction of the anomalous magnetic moment of the muon of
\beq \label{amuSMfinal}
a_{\mu}^{\rm SM}  =  (11\ 659 \ 181.08  \pm 3.78) \times 10^{-10} \, ,
\eeq
where the uncertainty is determined from the uncertainties of the individual SM contributions added in quadrature. This value deviates from the current experimental measurement~\cite{Bennett:2006fi} by
\begin{align} 
\Delta a_{\mu} = (28.02 \pm 7.37)\times 10^{-10}\, ,
\end{align}
corresponding to a muon $g-2$ discrepancy of $3.8\sigma$. This result
is compared with other determinations of $a_{\mu}^{\rm SM}$ in
Figure~\ref{amuCompare}. The value for $a_{\mu}^{\rm SM}$ in
equation~\eqref{amuSMfinal} has decreased by $0.96\times10^{-10}$
compared to the KNT18 analysis~\cite{Keshavarzi:2018mgv}. This 
change comes, in nearly equal parts, from the reduction in the mean 
value of $a_{\mu}^{\rm had, \, LO \, VP}$ and the new estimate of
 $a_{\mu}^{\rm had, \, LbL}$ in this work. The increase in the uncertainty
 with respect to~\cite{Keshavarzi:2018mgv} comes from the increase 
in the error of $a_{\mu}^{\rm had, \, LbL}$ owing 
to the changes in the estimate of this contribution discussed previously. 
Together, these have resulted in the increased discrepancy from $3.7\sigma$ 
in the KNT18 analysis to $3.8\sigma$ in this work.

\subsection{The anomalous magnetic moment of the tau lepton, $a_\tau$}\label{sec:a_tau}

In the case of the $\tau$, the determination of the LO hadronic VP contributions yields
\begin{align} \label{atauLOHVP_KNT19}
a_{\tau}^{\rm had, \, LO \, VP} & = (332.81  \pm 0.47_{\rm stat} \pm 1.09_{\rm sys} \pm 0.17_{\rm vp} \pm 0.69_{\rm fsr}) \times 10^{-8}  \nonumber
\\
\
& =  (332.81 \pm 1.39_{\rm tot}) \times 10^{-8} \ , 
\end{align}
whilst at NLO they are found to be
\begin{align}\label{atauNLOHVP_KNT19}
a_{\tau}^{\rm had,\, NLO\, VP} & = (7.85 \pm 0.01_{\rm stat} \pm 0.03_{\rm sys} \pm 0.01_{\rm vp} \pm 0.02_{\rm fsr}) \times 10^{-8}  \nonumber
\\
\
& =  (7.85 \pm 0.04_{\rm tot}) \times 10^{-8} \ .
\end{align}
Note that in the case of the $\tau$, the total NLO contributions are positive,
while they are negative for the electron and muon, and any estimate
based on a naive mass-scaling of the result for the muon would fail completely. 
The results for $a_{\tau}^{\rm had, \, LO \, VP}$ from the individual
hadronic channels are given in
Table~\ref{tab:comparisonTable}. Comparing with the evaluation in~\cite{Eidelman:2007sb}, which resulted in $a_{\tau}^{\rm had, \,
  LO \, VP} = (337.5 \pm 3.7) \times 10^{-8}$, and $a_{\tau}^{\rm had, \, NLO \, VP} =
(7.6 \pm 0.2) \times 10^{-8}$ obtained already in~\cite{Krause:1996rf}, there is
consistency between the mean values found in the different
analyses. However, there is a large reduction in the error in this
work which is mainly due to the abundance of precise new data
since~\cite{Eidelman:2007sb}. Utilising the values
from~\cite{Eidelman:2007sb} for the QED, EW and hadronic LbL
contributions (listed in Table~\ref{tab:atauSM-KNT19}), the updates to
the hadronic VP contributions result in a SM prediction for the
anomalous magnetic moment of the tau lepton of
\beq\label{atauSM}
a_{\tau}^{\rm SM} = (117717.1 \pm 3.9) \times 10^{-8}  \, .
\eeq
With the uncertainties of the hadronic VP contributions significantly
improved, the uncertainty of $a_{\tau}^{\rm SM}$ is now dominated by
the hadronic LbL contributions, which account for $\sim 60\%$ of the
total error. However, it should be noted that the QED contributions,
at $\sim 26\%$ of the total error, are now less precise than the
hadronic VP contributions. As explained in~\cite{Eidelman:2007sb}, the
entire error $\delta a_{\tau}^{\rm QED} \sim 2 \times 10^{-8}$ is
assigned as the uncertainty due to the missing contributions at
four-loop (and beyond), and are crudely estimated from logarithmically
enhanced terms expected at four-loop level. This indicates that a
calculation of $a_{\tau}^{\rm QED}$ at four loops would significantly
improve the determination of $a_\tau^{\rm SM}$.

Although, as stated in Section~\ref{sec:Intro}, the precision of the
current experimental measurement of $a_{\tau}^{\rm exp} =
-0.018(17)$~\cite{Abdallah:2003xd} makes a meaningful comparison
between theory and experiment futile, this analysis confirms a
difference $\Delta a_{\tau} = a_{\tau}^{\rm exp}-a_{\tau}^{\rm SM}$ at
the level of $1\sigma$ as found in~\cite{Abdallah:2003xd}. 
While at present there seems little prospect for an experiment
dedicated to measuring $a_\tau$, it is not imperceivable to imagine
that this might become possible in the future. Indeed, the additional
potential for new physics discoveries due to the higher mass scale of
the $\tau$ compared to the electron or the muon make this an interesting
consideration.

\begin{table}[!t]
\centering
\scalebox{1.0}{
{\renewcommand{\arraystretch}{1.1}
\begin{tabular}{|l|c|}
\hline 
{\bf SM contribution} &  $a_{\tau}$  \\
\hline 
QED & $ (117324.0 \pm 2.0) \times 10^{-8} $~\cite{Eidelman:2007sb} \\
EW & $ \hphantom{1173} (47.4 \pm 0.5) \times 10^{-8} $~\cite{Eidelman:2007sb} \\
had LO VP & $ \hphantom{117} (332.8 \pm 1.4) \times 10^{-8} $~\hphantom{[12]}   \\
had NLO VP & $ \hphantom{11732}  (7.9 \pm 0.0) \times 10^{-8} $~\hphantom{[12]}   \\
had LbL & $ \hphantom{11732} (5.0 \pm 3.0) \times 10^{-8}   $~\cite{Eidelman:2007sb}   \\
\hline
Theory total & $ (117717.1 \pm 3.9) \times 10^{-8} $~\hphantom{[12]}  \\ 
Experiment & $ \hphantom{1173} \hphantom{(}-0.018 \pm 0.017 \hphantom{\times 10^{-8}}$~\cite{Abdallah:2003xd} \hphantom{,)}  \hphantom{-}  \\ 
\hline 
$\Delta a_{\tau}$ & $ \ -0.019 \pm 0.017 \ (-1.1\sigma)  $  \\ 
\hline 
\end{tabular} 
}
}\caption{Summary of the contributions to $a_{\tau}^{\rm SM}$.}\label{tab:atauSM-KNT19} 
\end{table}

\subsection{Determination of $\alpha(M_{Z}^2)$}\label{sec:alphaMz}

The running (scale dependent) QED coupling, $\alpha(q^2)$, is
determined via \allowbreak $\alpha(q^2)=\alpha/\big(1-\Delta\alpha_{\rm
  had}(q^2)$\allowbreak$- \Delta\alpha_{\rm lep}(q^2)\big)$, where the contributions
to the running are separated into hadronic (had) and leptonic (lep)
components. Of the three fundamental EW parameters of the SM (the
Fermi constant $G_F$, $M_Z$ and $\alpha(M_{Z}^2)$), the effective QED
coupling 
at the $Z$ boson mass, $\alpha(M_{Z}^2)$, is the least precisely
known, where the uncertainties from the non-perturbative, hadronic
contributions limit the accuracy of EW precision fits. The
five-flavour (all quark flavours except the top quark which can be
treated perturbatively) contributions to $\alpha(M_{Z}^2)$ are
determined from the dispersion relation 
\beq \label{eq:delAlpha}
\Delta\alpha_{\rm had}^{(5)}(M_{Z}^2) = -\frac{\alpha M_{Z}^2}{3\pi}\,{\rm P}
\int^{\infty}_{s_{th}} {\rm d}s\frac{R(s)}{s(s-M_{Z}^2)}\,,
 \eeq
where P indicates the principal value of the integral.
Using the updated compilation for $R(s)$ from this work, and
perturbative QCD for energies $\sqrt{s}> 11.199$\ GeV (above the
thresholds for all five quark flavours), this data-driven evaluation
gives the result 
\begin{align} \label{delalphahad_KNT19}
\Delta\alpha_{\rm had}^{(5)}(M_Z^2) & = (276.09  \pm  0.26_{\rm stat} \pm 0.68_{\rm sys} \pm 0.14_{\rm vp} \pm 0.83_{\rm fsr}) \times 10^{-4}  \nonumber
\\
\
& = (276.09 \pm 1.12_{\rm tot}) \times 10^{-4} \ .
\end{align}
From this, the total value of the QED coupling at the Z boson mass is
\begin{align}
\alpha^{-1}(M_Z^2) & = \Big(1-\Delta\alpha_{\rm lep}(M_Z^2)-\Delta\alpha_{\rm had}^{(5)}(M_Z^2)-\Delta\alpha_{\rm top}(M_Z^2)\Big)\alpha^{-1} \nonumber
\\
\
 & = 128.946 \pm 0.015\, ,
\end{align}
updating the result from~\cite{Keshavarzi:2018mgv}. As
in~\cite{Keshavarzi:2018mgv}, the leptonic contribution is 
$\Delta\alpha_{\rm lep}(M_Z^2) = (314.979 \pm 0.002) \times
10^{-4}$~\cite{Steinhauser:1998rq,Sturm:2013uka}. The contribution
from the top quark is updated from~\cite{Chetyrkin:1995ii,Kuhn:1998ze}
by using $m_t = 172.9 (0.4)$ GeV,
$\alpha_s(M_Z)=0.1181(11)$~\cite{PDG2018} and by including the
contributions from ${\cal O}(\alpha_s^0 m_Z^6/m_t^6)$ and ${\cal
  O}(\alpha_s^1 m_Z^6/m_t^6)$ terms which were neglected 
in~\cite{Kuhn:1998ze}. This results in $\Delta\alpha_{\rm
  top}(M_Z^2) = (-0.7201 \pm 0.0037) \times 10^{-4}$.
A comparison with previous, largely data-driven determinations of
$\Delta\alpha_{\rm had}^{(5)}(M_Z^2)$ and $\alpha^{-1}(M_Z^2)$ is
given in Table~\ref{tab:comapredelalpha}.
\begin{table}[!t]
\vspace{-0.cm}
\centering
\scalebox{1.0}{
  {\renewcommand{\arraystretch}{1.0}
  \begin{tabular}{|l|c|c|}
\hline													
{\bf Analysis}	&	$\Delta\alpha_{\rm had}^{(5)}(M_Z^2)\times10^{4}$			&	$\alpha^{-1}(M_Z^2)$	\\
\hline	
DHMZ10~\cite{Davier:2010nc}	&	$	275.59 \pm 1.04	$	&	$	128.952 \pm 0.014 	$	\\			
HLMNT11~\cite{Hagiwara:2011af} &	$	276.26 \pm 1.38	$	&	$	128.944 \pm 0.019	$	\\
FJ17~\cite{Jegerlehner:2017zsb}	&	$	277.38 \pm 1.19	$	&	$	 128.919 \pm 0.022	$	\\
DHMZ17~\cite{Davier:2017zfy}	&	$	276.00 \pm 0.94	$	&	$	128.947 \pm 0.012	$	\\			
KNT18 	&	$	276.11 \pm 1.11	$	&	$	128.946 \pm 0.015	$	\\
DHMZ19~\cite{Davier:2019can}	&	$	276.10 \pm 1.00	$	&	$	128.946 \pm 0.013	$	\\
KNT19 [This work]	&	$	276.09 \pm 1.12	$	&	$	128.946 \pm 0.015	$	\\
\hline													
  \end{tabular} 
  }
}\caption{Comparison of recent and previous evaluations of $\Delta\alpha_{\rm had}^{(5)}(M_Z^2)$ determined from $e^+e^-\rightarrow {\rm hadrons}$ cross section data and the corresponding results for $\alpha^{-1}(M_Z^2)$.}\label{tab:comapredelalpha} 
\end{table}

\subsection{The hyperfine splitting of muonium, $\Delta\nu_{\rm Mu}^{\rm had, \, VP}$}\label{sec:hypMuon}

For many years, precision measurements of the ground-state hyperfine splitting (HFS) of muonium $\Delta\nu_{\rm Mu}$ served as a rigorous test of QED. Today, it still provides the best approach for determining the value of the electron-to-muon mass ratio and, therefore, the muon mass. As, like with the lepton $g-2$, $\Delta\nu_{\rm Mu}$ is sensitive to quantum effects, any differences in the comparison of experimental and theoretical determinations could be an indication of new physics. The current most precise experimental measurements of $\Delta\nu_{\rm Mu}$~\cite{Mariam:1982bq,Liu:1999iz} result in
\beq\label{eq:HFS_exp}
\Delta\nu_{\rm Mu}^{\rm exp} = (4 \ 463 \ 302 \ 776 \pm 51) \ {\rm Hz} \, .
\eeq
With the most recent of these measurements having been performed more than 20 years ago, the MuSEUM experiment at J-PARC is currently in the process of measuring the HFS of muonium (and the electron-to-muon mass ratio) with an aim to reduce the uncertainty in equation~\eqref{eq:HFS_exp} by an order of magnitude~\cite{MuSEUM}. 

The theoretical prediction, $\Delta\nu_{\rm Mu}^{\rm SM}$, as given by
CODATA 2014~\cite{Mohr:2015ccw}\footnote{Note that in~\cite{Eides:2018rph}
 it was claimed that the uncertainty in
  equation~\eqref{eq:HFS_SM_CODATA} is underestimated by a factor of $\sim
  1/2$ due to the implicit assumption that there is no new physics
  beyond the SM in relations used by the CODATA estimate. 
The theoretical (th) prediction in~\cite{Eides:2018rph} reads $\Delta\nu_{\rm Mu}^{\rm th} = 
(4 \ 463 \ 302 \ 872 \pm 515) \ {\rm Hz}$.}, is
\beq\label{eq:HFS_SM_CODATA}
\Delta\nu_{\rm Mu}^{\rm SM}({\rm CODATA}) = (4 \ 463 \ 302 \ 868 \pm
271) \ {\rm Hz} \,.
\eeq
Although the HFS of muonium is mainly QED dominated, it receives higher-order contributions from the EW and hadronic sectors. In the case of the hadronic contributions, the hadronic LO VP contributions are dominant, whilst the hadronic LbL contributions are negligible compared to the current level of precision ($\Delta\nu_{\rm Mu}^{\rm had, \, LbL} \simeq 0.0065(10) \ {\rm Hz}$~\cite{Karshenboim:2008gg,Mohr:2015ccw}). The CODATA determination given in equation~\eqref{eq:HFS_SM_CODATA} currently utilises the value for the hadronic LO VP contributions that was determined in the NT12 analysis preceding this work~\cite{Nomura:2012sb}, which found
\beq \label{eq:HFS_HVP_NT12}
\Delta\nu_{\rm Mu}^{\rm had, \, VP}({\rm NT}12) = (232.68 \pm 1.44) \ {\rm Hz} \, .
\eeq
These contributions can be determined via the dispersion integral
\beq\label{eq:HFShadVP}
\Delta\nu_{\rm Mu}^{\rm had, \, VP} = \frac{1}{2\pi^3}\frac{m_e}{m_\mu}\nu_F\int^{\infty}_{m_{\pi^0}^2} \ {\rm d}s \, K_{\rm Mu}(s)\sigma^0_{{\rm had},\gamma} (s) \, .
\eeq
Here, $\nu_F$ denotes the so-called Fermi energy,
\beq
\nu_F = \frac{16}{3}R_\infty\alpha^2\frac{m_e}{m_\mu}\left[1+\frac{m_e}{m_\mu}\right]^{-3} \, ,
\eeq
where $R_\infty$ is the Rydberg constant. The kernel function $K_{\rm Mu}(s)$ is described in detail in~\cite{Nomura:2012sb}. 

Now utilising the compilation of the hadronic cross section determined in this work (see Section~\ref{sec:DataUpdates}), the updated value for the hadronic VP contributions to the ground-state HFS of muonium are found to be
\begin{align}\label{eq:HFS_HVP_KNT19}
\Delta\nu_{\rm Mu}^{\rm had, \, VP} & = (232.04  \pm 0.38_{\rm stat} \pm 0.66_{\rm sys} \pm 0.08_{\rm vp} \pm 0.27_{\rm fsr}) \ {\rm Hz}  \nonumber
\\
\
& = (232.04 \pm 0.82_{\rm tot}) \ {\rm Hz} \ .
\end{align}
Here, a noticeable mean value reduction and an uncertainty reduction of $\sim 43\%$ compared to equation~\eqref{eq:HFS_HVP_NT12} are observed, which is in accordance with the same trends seen in the development of the corresponding determinations of $a_\mu$ over the same period. Adjusting the theoretical prediction in equation~\eqref{eq:HFS_SM_CODATA} for this value results in
\beq\label{eq:HFS_SM_KNT19}
\Delta\nu_{\rm Mu}^{\rm SM} = (4 \ 463 \ 302 \ 867 \pm 271) \ {\rm Hz} \,, 
\eeq
which, despite the noticeable changes in $\Delta\nu_{\rm Mu}^{\rm had, \, VP}$ between this work and the previous analysis, highlights the minimal impact of the hadronic contributions to this observable compared to the dominant QED contributions.

\section{Conclusions and future prospects}\label{Conclusions}

This analysis, KNT19, has presented updated evaluations of the hadronic
vacuum polarisation contributions to the anomalous magnetic moment of
the electron ($a_{e}^{\rm had, \, VP}$), muon ($a_{\mu}^{\rm had, \, VP}$) 
and tau lepton ($a_{\tau}^{\rm had, \, VP}$), to the ground-state hyperfine 
splitting of muonium ($\Delta\nu_{\rm Mu}^{\rm had, \, VP}$), and has also 
updated the value of the hadronic contributions to the running of the QED 
coupling at the scale of the mass of the $Z$ boson ($\Delta\alpha_{\rm had}(M_Z^2)$). 
These quantities are calculated using the hadronic $R$-ratio, obtained
from a compilation of all available $e^+e^- \rightarrow {\rm hadrons}$ 
cross section data. In this work, the data compilation has been
updated from the determination in~\cite{Keshavarzi:2018mgv},
accounting for new measurements. In the dominant $\pi^+\pi^-$ channel,
the inclusion of the CLEO-c data~\cite{Xiao:2017dqv} has increased the
mean value slightly and marginally improved the uncertainty of
$a_{\mu}^{\pi^+\pi^-}$. In the $\pi^+\pi^-\pi^0$ channel, adjustments
have been made to the treatment of the narrow $\omega$ resonance,
which is now integrated over using a quintic polynomial interpolation 
in order to avoid an overestimation of the cross section from a linear 
interpolation that was recently noted in~\cite{Hoferichter:2019gzf}.
This has reduced the mean value of $a_{\mu}^{\pi^+\pi^-\pi^0}$ by 
$\sim 1\times 10^{-10}$ and, in turn, contributed to a significant
reduction of the mean value of $a_{\mu}^{\rm SM}$ in this work,
although it is important to note that all estimates from this analysis
are consistent with those given in~\cite{Keshavarzi:2018mgv}. In
addition, other new measurements have been included which have removed
the need to rely on isospin relations to estimate cross sections in
three sub-leading channels, where in each case the new data agree well
with the predictions of the KNT18 analysis.

The resulting hadronic $R$-ratio has been used as input into
dispersion relations to determine $a_{l}^{\rm had, \, VP}$
($l=e,\mu,\tau)$ at LO and NLO, $\Delta\alpha_{\rm had}(M_Z^2)$ and
$\Delta\nu_{\rm Mu}^{\rm had, \, VP}$. This work has found
$\Delta\alpha_{\rm had}^{(5)}(M_Z^2) = (276.09 \pm 1.12_{\rm tot})
\times 10^{-4}$ which has yielded a value for the QED coupling at the
$Z$ boson mass of $\alpha^{-1}(M_Z^2) = 128.946 \pm 0.015$, which is
consistent with~\cite{Keshavarzi:2018mgv}. For the hadronic VP
contributions to the ground-state hyperfine splitting of muonium, the
new data compilation gives $\Delta\nu_{\rm Mu}^{\rm had, \, VP} =
(232.04 \pm 0.82_{\rm tot}) \ {\rm Hz}$, which is consistent with the
previous determination of this quantity in~\cite{Nomura:2012sb}, but
constitutes a significant uncertainty reduction of $\sim 43\%$. A
similar error reduction has been observed in the determination of the
anomalous magnetic moment of the electron compared
to~\cite{Nomura:2012sb}, with this analysis finding $a_{e}^{\rm had,
  \, LO \, VP} = (186.08 \pm 0.66_{\rm tot}) \times 10^{-14}$. This,
coupled with new estimates for the NLO contributions, translates to
differences between experiment and theory of $\Delta a_{e}(\alpha_{\rm
  Rb}) = (-1.312 \pm 0.773) \times 10^{-12}$ ($1.7\sigma$) and $\Delta
a_{e}(\alpha_{\rm Cs}) =  (-0.890 \pm 0.362) \times 10^{-12}$
($2.5\sigma$), depending on whether the QED contributions are
determined using $\alpha$ measured via Rb or Cs atomic
interferometry. For the muon $g-2$, the new KNT19 analysis gives 
$a_{\mu}^{\rm had, \, LO \, VP} =  (692.78 \pm 2.42_{\rm tot}) \times
10^{-10}$ and $a_{\mu}^{\rm had,\, NLO\, VP} =  (-9.83 \pm 0.04_{\rm
  tot}) \times 10^{-10}$. New choices in this work for the hadronic
LbL contributions based on recent results from dispersive approaches
(which have already significantly consolidated the `Glasgow consensus'),
coupled with the contributions from the other sectors of the SM, have
resulted in a new estimate for the Standard Model prediction of
$a_{\mu}^{\rm SM}  =  (11\ 659 \ 181.08  \pm 3.78) \times 10^{-10}$,
which deviates from the current experimental measurement by
$3.8\sigma$. In the case of the $\tau$, the value at LO is
$a_{\tau}^{\rm had, \, LO \, VP} =  (332.81 \pm 1.39_{\rm tot}) \times
10^{-8}$, consistent with the value found in~\cite{Eidelman:2007sb},
but with an uncertainty that is smaller by $\sim 62\%$. Unfortunately,
the current experimental bounds of the measured value of $a_\tau$ are
not stringent enough to draw any strong conclusions from the
comparison between experiment and theory.

It is interesting to compare the values and uncertainties of
$a_{l}^{\rm had, \, LO \, VP}$ and $a_{l}^{\rm SM}$ of the different
leptons, which are shown in Table~\ref{tab:alCompare}. Here,
especially in the case of the hadronic contributions, the difference
in the resulting magnitudes of these values due to lepton mass-scaling
arguments is evident. Indeed, in the most extreme example, the value
of $a_{l}^{\rm had, \, LO \, VP}$ is $\mathcal{O}(10^6)$ times larger
for the $\tau$ than for the electron. For $a_{l}^{\rm SM}$, the most
striking difference is in the level of the precision between the
different leptons. The electron, being less sensitive to hadronic
effects than the muon or the $\tau$, is by far the most
precise. However, the larger uncertainty of $a_{\tau}^{\rm SM}$
compared to $a_{\mu}^{\rm SM}$ is not solely due to hadronic
contributions (where, for the muon, the hadronic LbL estimates are
more accurate than for the $\tau$). Instead, as noted in
Section~\ref{sec:a_tau}, the uncertainty assigned due to the missing
four-loop contributions is a main cause of this disparity and could be
improved through the calculation of $a_{\tau}^{\rm QED}$ at four-loop
order.
\begin{table}[!t]
\centering
\addtolength{\tabcolsep}{0.1cm}
{\renewcommand{\arraystretch}{1.2}
\begin{tabular}{| c | S | S |}
\hline
Lepton flavour, $l$ & $a_{l}^{\rm had, \, LO \, VP} \times {10}^{7}$ &
                                                                       $a_{l}^{\rm SM} \times {10}^{7}$ \\ 
\hline
$e$ & \hspace{-0.01cm} 0.00001861(7) & \hspace{-0.01cm}11596.52182042(720) \\
$\mu$ & \hspace{-0.01cm}  0.69278(242) &  \hspace{-0.01cm}11659.18108(378) \\
$\tau$ & \hspace{-0.18cm} 33.281(139) & \hspace{-0.01cm}11771.71(39) \\
\hline
\end{tabular}
}\caption{Comparison of the contributions to $a_{l}^{\rm had, \, LO \, VP}$ and $a_{l}^{\rm SM}$ as determined in this work. All results are presented in units of $a_{l}\times10^7$ in order to compare the relevant magnitudes and precision of the various contributions. In this instance, the value of $a_{e}^{\rm SM}$ corresponds to $a_{e}^{\rm QED}$ determined using $\alpha_{\rm Rb}$.}\label{tab:alCompare} 
\end{table}

With the tantalising prospect of new experimental measurements of
$a_\mu$ from Fermilab in the near future, and later from J-PARC, the
predictions of $a_{\mu}^{\rm had, \, VP}$ and $a_{\mu}^{\rm SM}$ have
been re-examined in detail and found to be robust. The opportunity to
further improve the hadronic VP contributions estimated by dispersive
approaches (as in this analysis) largely rests on new hadronic cross
section measurements. For the $\pi^+\pi^-$ channel, new measurements
currently under analysis from the CMD-3, SND and BaBar experiments are
eagerly awaited. Although these measurements are important in terms of
improving the overall precision of $a_{\mu}^{\rm had, \, VP}$, it is
hoped that they will help to resolve the lingering deviation between
the
KLOE~\cite{Ambrosino:2008aa,Ambrosino:2010bv,Babusci:2012rp,KLOEcombination}
and BaBar~\cite{Aubert:2009ad} measurements, which drive the data tensions in 
$a_{\mu}^{\pi^+\pi^-}$. In addition, expected data for the
$\pi^+\pi^-\pi^0$,  $\pi^+\pi^-\pi^0\pi^0$ and the inclusive channels,
will be very beneficial. In preparation for the new experimental
measurements of $a_\mu$, the efforts of the Muon $g-2$ Theory
Initiative~\cite{TGm2} (and the groups within it) have already led to
impressive achievements with regards to advancing the determinations
of the hadronic VP and hadronic LbL contributions. Of great interest
are the results from lattice QCD, which already provide first-principles
cross checks of the now very precise data-driven estimates for the
hadronic contributions to $a_\mu^{\rm SM}$. These are expected to
become competitive with the current determinations within the next few
years. Given the continued advancements in the theoretical predictions
of $a_\mu$, coupled with the substantial progress of the experimental
community, the study of the muon anomalous magnetic moment has never been better
placed to severely constrain many scenarios for new physics beyond the
SM, or, should the muon $g-2$ discrepancy become fully established, to
claim a discovery of new physics.

\section*{Acknowledgements}

We would like to thank Martin Hoferichter, Bai-Long Hoid, Bastian
Kubis, the DHMZ group (Michel Davier, Andreas Hoecker, Bogdan Malaescu
and Zhiqing Zhang) and, in general, {\em The Muon $g-2$ Theory
  Initiative} and the Muon $g-2$ collaboration for numerous useful
discussions. Alex Keshavarzi would like to thank Tsutomu
Mibe and the KEK Laboratory for hosting him during part of the writing
of this paper.

The work of Alex Keshavarzi is supported in-part by STFC under the
consolidated grant ST/S000925/1. This manuscript has been authored by
an employee of The University of Mississippi (A.K.), supported in-part
by the U.S. Department of Energy Office of Science, Office of High
Energy Physics, award DE-SC0012391. This document was prepared using
the resources of the Fermi National Accelerator Laboratory (Fermilab),
a U.S. Department of Energy, Office of Science, HEP User
Facility. Fermilab is managed by Fermi Research Alliance, LLC (FRA),
acting under Contract No. DE-AC02-07CH11359. The work of Daisuke
Nomura is supported by JSPS KAKENHI grant number JP17H01133. The work
of Thomas Teubner is supported by STFC under the consolidated grants
ST/P000290/1 and ST/S000879/1.% old PP CG: ST/N000331/1

\end{document}